\documentclass[twocolumn,showpacs,preprintnumbers,floatfix,pra,aps]{revtex4-1}

\usepackage{graphicx}
\usepackage{dcolumn}
\usepackage{amsmath}
\usepackage{bbm}
\usepackage{hyperref}
\usepackage{feynmp}
\usepackage{color}

\renewcommand{\vec}[1]{{\mathbf #1}}
\newcommand{\ttt}{T_0}

\begin{document}
\title{The Floquet-Boltzmann equation}
\author{Maximilian Genske$^1$}
\email{genske@thp.uni-koeln.de}
\author{Achim Rosch$^1$}
\affiliation{$^1$ Institut f\"ur Theoretische Physik, Universit\"at zu K\"oln, D-50937 Cologne, Germany\\
}
\date{\today}

\begin{abstract}
Periodically driven quantum systems can be used to realize quantum pumps, ratchets, artificial gauge fields and novel topological states of matter.
Starting from the Keldysh approach, we develop a formalism, the Floquet-Boltzmann equation, to describe the dynamics and the scattering of quasiparticles in such systems. 
The theory builds on a separation of time-scales. Rapid, periodic oscillations occurring on a time scale $\ttt=2 \pi/\Omega$, are treated using the Floquet formalism and quasiparticles
are defined as eigenstates of a non-interacting Floquet Hamiltonian. The dynamics on much longer time scales, however, is modelled by a Boltzmann equation which describes the semiclassical dynamics
of the Floquet-quasiparticles and their scattering processes. As the energy is conserved only modulo $\hbar \Omega$, the interacting system heats up in the long-time limit.
As a first application of this approach, we compute the heating rate for a cold-atom system, where a periodical shaking of the lattice was used to realize the Haldane model \cite{Jotzu2014}. 
\end{abstract}

\maketitle
Periodically modulated quantum systems can effectively be described by a static Hamiltonian. 
This theoretical concept has recently evolved into a  major experimental tool used by many groups to generate new states of matter.

Early experiments \cite{Lignier2007,Struck2011} used, for example, that one can effectively change the strength as well as sign of the hopping of atoms in an optical lattice, allowing to realize new types of band structures. 
Periodic driving has also be used to realize directed transport in quantum ratchets \cite{Salger2009}.
More recently, the realization of emergent Gauge fields and topological band structures has been at the focus of many studies. 
Examples of such experiments include the generation of Gauge fields and superfluids with finite momentum\cite{Struck2012,Struck2013}, 
the generation of topological quantum walks \cite{Kitagawa2012} and of effective electric fields in a discrete quantum simulator \cite{Genske2013},
the realization of (Floquet-) topological insulators with photons \cite{Rechtsman2013},
 the generation of spin-orbit coupling \cite{Anderson2013}, the direct measurement of Chern numbers and Berry phases in the Hofstadter Hamiltonian \cite{Aidelsburger2013,Aidelsburger2015} 
 and the realization of quantum pumps \cite{Lohse2015}. 
A recent experiment of the Esslinger group beautifully realized the Haldane model \cite{Jotzu2014}, 
i.e., a model which demonstrates that a quantum-Hall state can exist without homogeneous external magnetic fields.  
In solid-state systems circularly polarized light has been used \cite{Wang2013} to manipulate the surface states of topological insulators.

Also from the theory side, many proposals have pointed out that periodically driven states can be used
to realized a wide range of states of matter.
Examples are photoinduced quantum Hall states \cite{Oka2009,Kitagawa2011} 
and various topological Floquet states \cite{Kitagawa2010,Lindner2011,Gomez2013,Rudner2013} 
including dissipative systems \cite{Dehghani2014}, 
quantum ratchets for Mott  insulators \cite{Genske2014}, 
Majorana Fermions in driven quantum wires \cite{Jiang2011}, 
the generation of non-abelian gauge fields \cite{Hauke2012},
Floquet fractional Chern insulators \cite{Grushin2014},  
Floquet-Anderson insulators and quantized charge pumps \cite{Titum2015}.

In a periodically driven system, the Hamiltonian has only a discrete time-translational symmetry, $H(t+\ttt)=H(t)$. 
As a consequence, the total energy is {\em not} conserved but quantized changes of energy are possible, $\Delta E =n\, \hbar \Omega$ with $\Omega=2 \pi/\ttt$ and $n\in \mathbbm Z$. 
For non-interacting systems the absence of energy conservation has mostly no effect. 
The situation is, however, different when interactions in a many-particle system are considered. 
For a generic closed system, one can expect that in the long-time limit, $t \to \infty$, the system approaches the state with the highest entropy consistent with the conservation laws. 
In the absence of some cooling mechanism, e.g., by an external bath \cite{Seetharam2015} or by emitting radiation, 
one can therefore expect that generic interacting Floquet systems heat up to infinite temperatures \cite{Eckstein2011} (an exception are many-body localized systems \cite{Ponte2015}). 
This important (and well-known) aspect has received relatively little attention in previous studies. Eckstein and Werner used time-dependent dynamical mean-field theory to study heating effects. 
In \cite{Choudhury2014}, the stability of BEC condensates in periodically driven systems discussed based on phase-space arguments for energy non-conserving scattering processes. 

The goal of this paper is to derive and apply a Floquet-Boltzmann equation, i.e., a kinetic equation which can be used to describe the dynamics of weakly interacting Floquet systems. 
Such a kinetic equation is perhaps the simplest theoretical description which captures microscopically how interactions can  equilibrate an interacting quantum system. 
Like other quasi-classical kinetic equations, our approach builds on a separation of time scales and describes situations, 
where the change of occupation functions is much slower than $\ttt$ and the time-scales set by the bare parameters of the Hamiltonian.
Motivated by the recent realization of the Haldane model by the Esslinger group \cite{Jotzu2014}, we investigate the effects of local interactions in a fermionic system.
Starting from the Keldysh-formalism, we obtain a  quantum kinetic equation from the Dyson equation,
which is then reduced to the Floquet-Boltzmann equation. 
As an example, we quantitatively investigate heating rates in a limit when energy-conserving processes dominate (realized in Ref. \cite{Jotzu2014}).

A similar semiclassical kinetic equation was also derived in a recent preprint of Seetharam {\it et al.} \cite{Seetharam2015}. 
In contrast to our work, they considered electron-phonon coupling instead of fermion-fermion interactions. 
Technically, Ref.~\cite{Seetharam2015} used an equation-of-motion approach. While this approach is, perhaps, less well suited to 
investigate limitations of semiclassics, we expect that in the semiclassical limit it gives results equivalent to our derivation.

\section{Keldysh approach and Quantum kinetic equation}

Our aim is to develop a Floquet-Boltzmann approach for interacting many-particle systems.
The derivation builds on two main elements: a separation of time scales and the limit of weak interactions.
We consider a time-dependent many-particle Hamiltonian $H(t)$ which has the property that on {\em short}, microscopic time scales it is approximately periodic,
\begin{eqnarray}H(t)\approx H(t+\ttt)
\end{eqnarray}
 with period $\ttt$. 
 Furthermore, we also allow for a slow time-dependence on time-scales large compared to the $\ttt$ and all microscopic time scales (inverse kinetic and potential energies and inverse scattering rates). 
 The latter can, for example, be used to describe the influence of external forces or the slow (quasi-adiabatic) change of the Hamiltonian. The Hamiltonian can therefore be written in the form
\begin{eqnarray}
H(t)= \sum_n H_n(t) e^{- i \Omega n t}
\end{eqnarray}
with $\Omega=\frac{2 \pi}{\ttt}$ and with Fourier series coefficients $H_n(t)=H_{-n}(t)^\dagger$ which are time-dependent  on time scales much {\em larger} than $\ttt$. 

As the derivation of quantum kinetic equations is ultimately based on perturbation theory, we further require that this perturbation theory can indeed be applied. 
In our examples, this will be the case when interactions are weak. 
More generally, one can use the approach also in cases where interactions are strong, 
but scattering rates are nevertheless weak either due to phase space restriction 
(e.g., for the fermionic quasiparticles of a Fermi liquid close to the Fermi surface) or just because the density of excitations is low (e.g., a weakly excited bosonic Mott insulator \cite{Genske2014}).


\begin{figure}[t]
    \centering
    \includegraphics{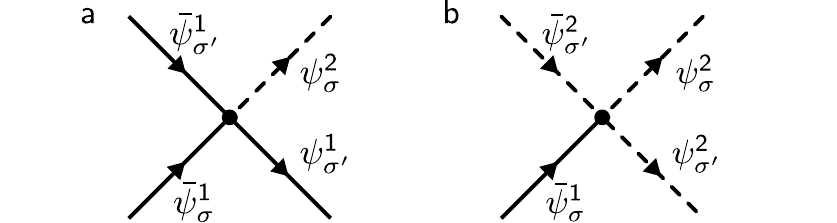}
    \caption{(Color online) Two of four interaction vertices generated by interaction of type Eq.~\eqref{HamInt}.
    The remaining two diagrams are obtained by complex conjugation of (a) and (b), which simply corresponds to inverting the direction of the arrows.}
    \label{vertices}
\end{figure}

To simplify the presentation, we will not discuss the most general setup but restrict ourselves to a simpler case.
We consider weakly interacting Fermions in a lattice model that is described by the following Hamiltonian
\begin{align}
 \label{Hamiltonian}
 \begin{split}
  H = H^0(t) + H^{\text{int}}
 \end{split}
\end{align}
with a static and local interaction
\begin{equation}
 \label{HamInt}
 H^{\text{int}} = U \sum_i c^{\dagger}_{i \uparrow} c^{\dagger}_{i \downarrow} c^{\phantom{\dagger}}_{i \downarrow} c^{\phantom{\dagger}}_{i \uparrow}
\end{equation}
We assume in the following that the interaction strength $U$ is sufficiently small, i.e., $U \ll J$, in order to allow a perturbative approach to solve the problem. 
Note that our approach can be generalized in a straightforward way to more complicated and time-dependent interactions.
We will present a theory that will unite the Floquet theory describing the oscillatory character of our model 
with the Keldysh approach of quantum field theories capturing the non-equilibrium behaviour of the system due to interactions and adiabatic drifts.

In order to derive a quantum kinetic equation, we use  the standard Keldysh approach \cite{Kamenev2011,Rammer1986}.  
We will not give a review of this Keldysh approach here, but just give a few main definitions. 
More details can, e.g., be found in the book by Kamenev\cite{Kamenev2011} on which the following discussion is based.

To describe the time-evolution of the density matrix, $\rho(t)=U(t) \rho_0 U(t)^\dagger$ one needs to keep track of two time-evolution operators $U(t)$ and $U(t)^\dagger$.
Within the functional-integral version of the Keldysh approach  one therefore introduces Grassmann fields $\psi$ on the forward branch, $\psi^{+}$, and on the backward branch, $\psi^{-}$.
One then performs a rotation in Keldysh space using the following relations
\begin{equation}
 \begin{pmatrix}
\psi^{1}\\
\psi^{2}
 \end{pmatrix}
 =
  \begin{pmatrix}
\frac{1}{\sqrt{2}}&\frac{1}{\sqrt{2}}\\
\frac{1}{\sqrt{2}}&\frac{-1}{\sqrt{2}}
 \end{pmatrix}
  \begin{pmatrix}
\psi^{+}\\
\psi^{-}
 \end{pmatrix}
 ,
  \begin{pmatrix}
\bar{\psi}^{1}\\
\bar{\psi}^{2}
 \end{pmatrix}
 =
  \begin{pmatrix}
\frac{1}{\sqrt{2}}&\frac{-1}{\sqrt{2}}\\
\frac{1}{\sqrt{2}}&\frac{1}{\sqrt{2}}
 \end{pmatrix}
  \begin{pmatrix}
\bar{\psi}^{+}\\
\bar{\psi}^{-}
 \end{pmatrix}
\end{equation}
After such a Keldysh rotation the action corresponding to the system excluding the interactions, i.e., $H-H_{\text{int}}$, can be written in the form
\begin{equation}
\label{S0}
S_{0} =
\int 
\begin{pmatrix}
 \bar{\psi}^{1} , \bar{\psi}^{2}
\end{pmatrix}
\hat{G}_0^{-1}
\begin{pmatrix}
 \psi^{1} \\
 \psi^{2}
\end{pmatrix}
\end{equation}
where the integral should be understood as $\int = \int dt \int dt' \sum_{ij}$, 
and the fields as $\bar{\psi}=\bar{\psi}_i(t)$ and $\psi=\psi_j(t')$. 
The non-interacting Green's function, $\hat{G}$, is a 2x2 matrix with components
 \begin{eqnarray}
\hat{G}_0^{-1}&=&\begin{pmatrix}
 (G^{-1}_0)^{R} & (G^{-1}_0)^{K} \\
  0 & (G^{-1}_0)^{A}
\end{pmatrix}
\end{eqnarray} 
Generally, the Green's functions of the full system are given by
\begin{align}
 G^R(x,x') &= -i \theta(t-t') \langle \{ \psi(x) , \psi^\dagger(x') \}  \rangle \nonumber \\
 G^A(x,x') &= i \theta(t'-t) \langle \{ \psi(x) , \psi^\dagger(x') \}  \rangle \nonumber \\
 G^K(x,x') &= - i \langle [ \psi(x) , \psi^\dagger(x') ]  \rangle \label{allGF}
\end{align} with $x= (\boldsymbol{r},t)$  
or within the functional integral by
\begin{equation}
\label{G}
G^{\alpha\beta}(x,x') =
- i \int D[\psi,\bar{\psi}]~
\psi^{\alpha}(x) \bar{\psi}^{\beta}(x')
e^{i (S_{0} + S_{\text{int}})}
\end{equation}
Note that we choose $\hbar=1$ throughout the paper. 
 Here, $G^{11}=G^{R}$ and $G^{22}=G^{A}$ are the retarded and advanced Green's functions, 
where $G^{R}=(G^{A})^{\dagger}$,
and $G^{12}=G^{K}$ is the Keldysh Green's function (note that $G^{21}=0$).
The latter can generally be parametrized in the following form
\begin{equation}
\label{GKpara}
G^{K} = G^{R} \circ F - F \circ G^{A}
\end{equation}
where $F=F(x,x')$ is a Hermitian matrix and called \textit{distribution matrix}.
The ``$\circ$'' represents matrix multiplication in all indices (space, time, spin).
While $G^R,G^A$ carry information about the spectrum of the system, $F$ holds the information about the occupation as we will see below.

The Dyson equation 
\begin{equation}
\label{Dyson}
( \hat{G}^{-1}_{0} - \hat{\Sigma}[\hat G] ) \circ \hat{G} = \hat{1}
\end{equation}
plays a central role in the derivation of the quantum kinetic equation. 
All objects are again matrices in Keldysh space, and the ``$\circ$'' now also includes a matrix multiplication with respect to Keldysh indices. 
Here, one uses that the self-energy,  $\hat{\Sigma}=\hat{\Sigma}[\hat G]$, can be viewed as a functional of the full Green's function, $\hat G$. 
The Dyson equation therefore is an integro-differental equation to determine $\hat G$ in a time dependent system. 
The Keldysh component of the equation above can be identified with the {\em quantum kinetic equation} (see below). 
Upon further approximations, this equation can be simplified to obtain the semiclassical Boltzmann equation. 

Using the fact that (in the fermionic case) the self-energy has the same structure as $\hat{G}^{-1}$ and $\hat{G}$, i.e.,
\begin{equation}
 \label{Sigma}
 \hat{\Sigma} =
 \begin{pmatrix}
 \Sigma^{R} & \Sigma^{K}\\ 
 0 & \Sigma^{A}
 \end{pmatrix}
\end{equation}
one can write down the Keldysh component of the Dyson equation, i.e. the \textit{quantum kinetic equation},
\begin{equation}
\label{QKE}
F\circ (G_{0}^{A})^{-1} - (G_{0}^{R})^{-1} \circ F = \Sigma^{K} - (\Sigma^{R} \circ F - F \circ \Sigma^{A})
\end{equation}

So far the expression is exact and no approximation has been made. 
For an interaction of the type as in Eq.\eqref{HamInt}, the interaction part of the action takes in Keldysh space the form 
\begin{eqnarray}
\label{Sint2}
 S_{\text{int}} =
- \frac{U}{2} \int \sum_{i,\sigma}
 && \bar{\psi}^{1}_{i \sigma}  \bar{\psi}^{1}_{i \bar\sigma} \psi^{1}_{i\bar\sigma} \psi^{2}_{i\sigma} \\[-5pt]
 && ~ + ~\bar{\psi}^{1}_{i\sigma} \bar{\psi}^{2}_{i\bar\sigma} \psi^{2}_{i\bar\sigma} \psi^{2}_{i\sigma} + \text{h.c.} \nonumber
\end{eqnarray}
where $\bar \sigma=\downarrow$ ($\uparrow$) for $\sigma=\uparrow$ ($\downarrow$).
Diagrammatically, $S_{\text{int}}$ yields four interaction vertices:
two independent ones plus their complex conjugates (see Fig.\ref{vertices}).  
To derive a quantum kinetic equation, we will consider (self-consistent) self-energy diagrams 
 up to second order.

To linear order in $U$, one obtains the familiar Hartree-Fock contributions: 
the energy of an $\uparrow$ electron is changed by $U \langle n_{i\downarrow}\rangle$. 
In the following, we will absorb all these completely into a redefinition of the non-interacting part $H^0$. 
Note, however, that due to the time dependence of expectation values $ \langle n_{i\downarrow}\rangle$, 
the non-interacting part will obtain an extra time-dependence which has to be calculated self-consistently.

%

\begin{figure}[t]
    \centering
    \includegraphics[width=8.5cm]{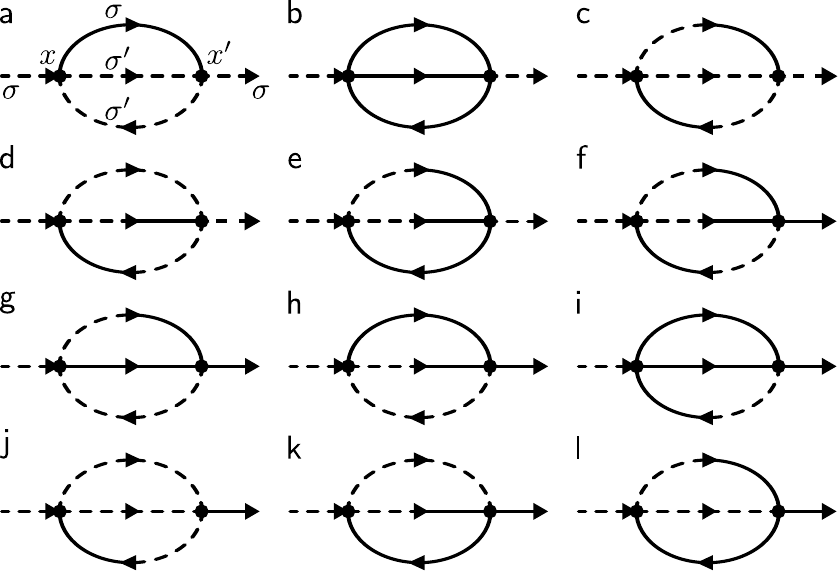}
    \caption{(Color online) Second order diagrams due to on-site interactions.
    (a)-(e) contribute to $\Sigma^{R}$ and (f)-(l) contribute to $\Sigma^{K}$.
    $\Sigma^{A}$ is obtained by realising that $\Sigma^{A}=\Sigma^{R\dagger}$.
    Note that the assignment of the spin indices as in (a) is the same for all diagrams.
    }
    \label{diagrams}
\end{figure}

Due to the fact that $G^{21}=0$, twelve independent diagrams are obtained in total for the second order expansion (see Fig.\ref{diagrams}) 
that contribute to the respective part of the self-energy.
Exploiting the general identities for retarded and advanced Green's functions $G^{R}(x',x)G^{A}(x,x')+G^{A}(x',x)G^{R}(x,x')=-G^{RA}(x',x)G^{RA}(x,x')$ 
and $(G^{A}(x,x'))^2+(G^{R}(x,x'))^2 = (G^{RA}(x,x'))^2$, where $G^{RA} = G^{R}-G^{A}$, one can write down the second order contributions to the individual parts of the self-energy as
\begin{align}
\label{SigmaK}
 \Sigma^{K}_{\sigma} &= - \frac{U^2}{4}
\Big[  G^{K}_{\bar\sigma}(x',x) ~ G^{K}_{\sigma}(x,x')~ G^{K}_{\bar\sigma}(x,x') \\
& \qquad\quad + ~G^{K}_{\bar\sigma}(x',x) ~ G^{RA}_{\sigma}(x,x') ~ G^{RA}_{\sigma}(x,x') \nonumber \\
& \qquad\quad - ~G^{RA}_{\bar\sigma}(x',x) ~ G^{K}_{\sigma}(x,x') ~ G^{RA}_{\bar\sigma}(x,x')  \nonumber\\
& \qquad\quad - ~G^{RA}_{\bar\sigma}(x',x) ~ G^{RA}_{\sigma}(x,x') ~ G^{K}_{\bar\sigma}(x,x') \Big]  \nonumber
\end{align}
\begin{align}
\label{SigmaR}
\Sigma^{R}_{\sigma} &= - \frac{U^2}{4}
\Big[  
G^{A}_{\bar\sigma}(x',x)~ G^{RA}_{\sigma}(x,x') ~G^{RA}_{\bar\sigma}(x,x') \\
& \qquad\quad + G^{A}_{\bar\sigma}(x',x) ~G^{K}_{\sigma}(x,x') ~G^{K}_{\bar\sigma}(x,x') \nonumber \\
& \qquad\quad + G^{K}_{\bar\sigma}(x',x) ~G^{K}_{\sigma}(x,x') ~G^{R}_{\bar\sigma}(x,x') \nonumber \\
& \qquad\quad + G^{K}_{\bar\sigma}(x',x) ~G^{R}_{\sigma}(x,x') ~G^{K}_{\bar\sigma}(x,x') \Big] \nonumber
\end{align}
where again $\bar \sigma=\downarrow$ ($\uparrow$) for $\sigma=\uparrow$ ($\downarrow$) is the conjugate spin to $\sigma$ and $\Sigma^{K}_{\uparrow}=\Sigma^{K}_{\uparrow\uparrow}(x,x')$.
The expression for $\Sigma^{A}$ can be straightforwardly obtained by using that $\Sigma^{A}=\Sigma^{R\dagger}$.

The equations for the self-energies (\ref{SigmaK},\ref{SigmaR}) as functions of the interacting Green's function together with the Dyson equation (\ref{Dyson}) 
and a suitable initial condition define a quantum kinetic equation (QKE), which can be used to study the dynamics of a time-dependent interacting system.

A direct (numerical) solution of the QKE is very challenging, especially for a time-dependent Hamiltonian
where the Green's function depends on two time-variables separately.
One can, however, make progress in situations where there is a clear separation of time scales.

The next step is to switch to a semi-classical representation via the Wigner transformation which we have to combine with the Floquet formalism to take into account  rapid periodic oscillations.

%
%
%
%

\section{Floquet eigenstates and Floquet Green's functions}
\label{sec:Floquet}
The analysis of the dynamics of the Floquet system starts from the non-interacting, but time-dependent part of the Hamiltonian,
$H^0(t)= \sum H^0_n(t) e^{- i \Omega n t}$, where  $H^0_n(t)$ varies only on time scales $\tau_{\rm slow} \gg \ttt$. 
More precisely, we will include in $H^0$ also all Hartree-Fock corrections arising from the interacting part of the Hamiltonian, 
i.e., terms like 
$U \sum_i c^\dagger_{i \uparrow}c^{\phantom{\dagger}}_{i \uparrow}  \langle n_{i\downarrow}(t)\rangle +c^\dagger_{i \downarrow}c^{\phantom{\dagger}}_{i \downarrow}  \langle n_{i\uparrow}(t)\rangle$ where $\langle n_{i\sigma}(t) \rangle$ 
will generically have components oscillating with frequencies $n \Omega$ plus an extra slow time-dependence arising, e.g., from heating.
Using the assumed separation of time scales allows to define  
\begin{eqnarray}
\label{Hamt0}
H^0_{t_0}(t)= \sum_{n,i,j} (h^0_{n}(t_0))_{ij}^{\phantom \dagger}  c^\dagger_i c^{\phantom \dagger}_j e^{- i \Omega n t}
\end{eqnarray}
as an approximation to $H^0(t)$ for $t$ close to $t_0$ (on the time scale set by $\tau_{\rm slow}$) with $H^0_{t_0}(t+\ttt)=H^0_{t_0}(t)$. 
To solve such exact time-periodic Hamiltonians one uses the Floquet theorem 
(the analog of the Bloch theorem for periodic time- instead of space-dependencies) 
stating that solutions can be written in the form $| \psi_{\nu}(t) \rangle = e^{-i\epsilon_\nu t}|\phi_{\nu}(t)\rangle $.
Here, $|\phi_{\nu}(t)\rangle=|\phi_{\nu}(t+\ttt)\rangle$ is a time-periodic Floquet state and $\epsilon_\nu$ is called the quasi energy.
The eigenstates $|\nu,t_0\rangle$ are obtained by diagonalizing the Floquet Hamiltonian
\begin{eqnarray}
\label{FloquetHam}
H^F_{t_0,nm}=n \Omega \delta_{nm}\, \mathbbm 1 + h^0_{n-m}(t_0)
\end{eqnarray}
with $H^F_{t_0} |\nu,t_0\rangle= \epsilon_{t_0 \nu}  |\nu,t_0\rangle$.
For practical calculations , the Floquet indices $n,m$ run from $-N_f$ to $N_f$, which is chosen such that $\Omega N_f$ is much larger than any other energy scale in the problem. 
Due to the `translation' invariance (obtained for  $N_f \to \infty$), $H^F_{t_0,n+1,n+1}=\Omega+H^F_{t_0,nn}$, 
one can obtain from each eigenstate with energy $\epsilon$ an eigenstate with energy $\epsilon + k \Omega$ by a simple translation of the Floquet indices, $n \to n+k$. 
Therefore, it is sufficient to consider only eigenstates with eigenenergies \begin{eqnarray}
-\frac{\Omega}{2} \le \epsilon_{t_0 \nu} < \frac{\Omega}{2}
\end{eqnarray}
Here, $\nu$ encodes the usual quantum number (spin, band-index, momentum).

The Heisenberg operator
\begin{align}
\label{FloquetHeisenberg}
f_{t_0, \nu}^\dagger(t) &=e^{-i \epsilon_{t_0, \nu} t} \sum_{i,n} \phi^n_{t_0,\nu}(i)  ~ e^{- i \Omega n t}  ~ c_i^\dagger
\end{align}
creates a fermion in such a Floquet eigenstate with $\phi^n_{t_0,\nu}(i)=\langle i,n|\nu,t_0\rangle$ (here $i$ includes all quantum numbers, e.g., lattice site and spin). 
Note that 
\begin{eqnarray}
\langle f_{t_0, \nu}^\dagger(t) 
 f_{t_0, \nu}^{\phantom{\dagger}}(t) \rangle=n_{t_0,\nu}
\end{eqnarray}
the occupation of the Floquet states,
 is {\em time independent} for the non-interacting, periodic Floquet Hamiltonian (\ref{Hamt0}). 

Here, it is important to stress that a single function $n_{t_0,\nu}$ describes the occupation of the Floquet states,
 and it is not necessary to introduce separate occupations for each Floquet index 
(further Floquet eigenstates obtained by translations in Floquet space yield exactly the same Floquet-creation operator, $f_{t_0, \nu}^\dagger(t)$).
  Our main goal will be to find a semiclassical description of the time evolution of the Floquet-occupations $n_{t,\nu}(t)$.

In a state with $\langle f_{t_0, \nu}^\dagger(t) 
 f^{\phantom{\dagger}}_{t_0, \nu'}(t) \rangle= \delta_{\nu \nu'} n_{t_0,\nu}$ one can
calculate the Green's functions (\ref{allGF}) of the noninteracting system (\ref{Hamt0}). 
They depend on two time variables but the dependence on $(t+t')/2$ is purely periodic and therefore it is useful to represent the Green's function in Floquet space by introducing the 
Floquet-Fourier transformation $A_{nm}(\omega)=\frac{1}{\ttt}\int dt\,dt' e^{i(\omega + n\Omega)t} e^{-i(\omega + m\Omega)t'} A(t,t')$ to obtain
\begin{align}
\label{GsFloquet} 
 G^R_{\substack{t_0,ij\\nm}}(\omega) &= \sum_{\nu,l}  \frac{\phi_{t_0,\nu}^{n+l}(i) ~ 
\bar{\phi}_{t_0,\nu}^{m+l}(j)}{\omega-\varepsilon_{t_0,\nu}-l \Omega+i0} \nonumber \\
G^A_{\substack{t_0,ij\\nm}}(\omega) &= \sum_{\nu,l}  \frac{\phi_{t_0,\nu}^{n+l}(i) ~ 
\bar{\phi}_{t_0,\nu}^{m+l}(j)}{\omega-\varepsilon_{t_0,\nu}-l \Omega-i0}  \\
\begin{split}
  G^K_{\substack{t_0,ij\\nm}}(\omega) &= - 2\pi i \sum_{\nu,l} \delta(\omega-\varepsilon_{t_0,\nu}-l \Omega) ~\times \\[-5pt]
&  \qquad \qquad \times ~\phi_{t_0,\nu}^{n+l}(i) ~ \bar{\phi}_{t_0,\nu}^{m+l}(j) ~ (1-2 n_{t_0,\nu}) 
\end{split} \nonumber
\end{align}
The distribution function defined in  (\ref{GKpara})  is therefore given in Floquet representation by
\begin{align}
 \label{FFloquet}
 F_{\substack{t_0,ij\\nm}}(\omega) = \sum_{\nu,l} \phi_{t_0,\nu}^{n+l}(i) ~ \bar{\phi}_{t_0,\nu}^{m+l}(j) ~ (1-2 n_{t_0,\nu})
\end{align}
Note that the Green's functions defined above are solutions of the Dyson equation 
\begin{eqnarray}
\sum_{m'} (\omega-\hat H_{t_0,nm'}^F) \hat G_{t_0,m' m}(\omega,t)=\delta_{nm} \hat{\mathbbm 1}
\label{dysonFM}
\end{eqnarray}

\section{Floquet-Wigner formalism and Floquet-Moyal expansion}
\label{Sec:FloquetWigner}
Our central goal is to separate the slow dynamics, treated within a semiclassical approach, from rapid, periodic oscillations which have to be treated fully quantum-mechanically. 
The latter aspect is treated within the Floquet formalism (see Sec.~\ref{sec:Floquet}) which uses that in a strictly periodic system with period $\ttt$, only Fourier modes of the form $n \Omega$ with $\Omega=\frac{2 \pi}{\ttt}$ occur. 
To derive the slow, semiclassical dynamics, the  starting point is the use of a Wigner representation of the Green's function $G(t,t')$,
usually obtained by using a Fourier transformation of the relative time coordinate, $t-t'$, for fixed $\bar t = \frac{t+t'}{2}$.

These two approaches can be combined in situations where there is a clear separation of time scales, as discussed above. 
We require that the oscillation period $\ttt$ is much smaller than all time scales, $\tau_{\rm slow}$, on which the occupation function changes or on which the oscillating Hamiltonian is modified.
This allows us to introduce the following `Floquet-Wigner representation' for functions $A(t,t')$ (Green's functions or self-energies) which depend on two time coordinates
\begin{eqnarray} \label{floquetWigner}
 A_{nm}(\omega,\bar t) =&& \\ \frac{1}{\ttt}\int dt\,dt' &&  \delta_\tau\!\!\left(\bar t-\frac{t+t'}{2}\right) ~ e^{i(\omega + n\Omega)t} e^{-i(\omega + m\Omega)t'} A(t,t')\nonumber
\end{eqnarray}
Here,
\begin{align}
 \delta_\tau\!\!\left(\bar t-t\right)  =\frac{1}{\sqrt{2 \pi} \tau} e^{-\left(\frac{\bar t -t}{\tau}\right)^2}
\end{align}
is a version of the $\delta$-function which is broadened on the time scale $\tau$ chosen much larger than the period $\ttt$, but much smaller than the time scale of slow modifications, $\tau_{\rm slow}$
\begin{eqnarray}
\ttt \ll \tau \ll \tau_{\rm slow} \label{separation}
\end{eqnarray}
The use of the filter function  $\delta_\tau\!\!\left(\bar t-\frac{t+t'}{2}\right)$ guarantees that $A_{nm}(\omega,\bar t)$ is {\em not} rapidly oscillating on the time scale $\ttt$. 
More precisely, all oscillating components at frequency $\Omega$ are {\em exponentially} suppressed by the factor 
$e^{-\left(\Omega \tau \right)^2/4}=e^{-\pi^2 \left(\tau/\ttt\right)^2}$ due to the convolution with the filter function.
Later, $\bar t$ will take over the role of $t_0$ introduced in the previous section.

The inverse transformation of the Floquet-Wigner representation is given by 
\begin{align}
\label{InversefloquetWigner}
 A(t,t') &\approx \sum_{n,m} \int \frac{d\omega}{2\pi} ~ e^{-i(\omega + n\Omega)t} e^{i(\omega + m\Omega)t'}~ A_{nm}(\omega,\frac{t+t'}{2})
\end{align}
Due to the finite width $\tau$ of the  filter function $\delta_\tau\!\!\left(t \right)$, 
this back-transformation is {\em not} exact but it is valid (with exponential precision) in situations when Eq.~(\ref{separation}) holds.
To see just this, it is instructive to plug Eq.~(\ref{InversefloquetWigner}) into  Eq.~(\ref{floquetWigner}): 
The condition $\ttt \ll \tau$ guarantees that the Floquet indices do not mix,
while $ \tau \ll \tau_{\rm slow}$  allows to use use $\delta_\tau(t)$ as a true $\delta$-function for all components which vary slowly in time.

By definition, the $\omega$ argument of the Floquet-Wigner representation is restricted to the interval $-\frac{\Omega}{2} \le \omega < \frac{\Omega}{2}$ 
(the analog of the reduced Brillouin zone for periodic systems in real space). 

Within the quantum kinetic equation (\ref{QKE}), one has to compute the product of matrices, $A = B \circ C$,
which takes the form
\begin{align}
A(t_2,t_1)= (B \circ C)(t_2,t_1)=\int dt' B(t_2,t') C(t',t_1)
\end{align}
Into this equation we plug the inverse Floquet-Wigner transformation, Eq.~(\ref{InversefloquetWigner}), 
for $B$ and $C$, and Taylor-expand the time and frequency arguments of $B_{nm}(\omega,t)$ and $C_{nm}(\omega,t)$. 
It turns out, that the resulting expression can be written in the form
\begin{equation}
 \label{MoyalFloquet}
 A_{nm}(\omega,t) = e^{-\frac{i}{2}(\partial_{t}^{B}\partial_{\omega}^{C} - \partial_{\omega}^{C}\partial_{t}^{B} )} \sum_{l} B_{nl}(\omega,t)C_{lm}(\omega,t) 
\end{equation}
Here $\partial_{t,\omega}^{A/B}$ denotes the time- or frequency derivative of the functions $A$ or $B$, respectively. 
By a Taylor-expansion of the exponential, one obtains the well-known Moyal expansion \cite{Kamenev2011}. 
The only difference in comparison to the standard Moyal expansion is that it is supplemented by a simple matrix multiplication of the Floquet indices. 
A short derivation of the Floquet-Moyal expansion can be found in appendix~\ref{appendixMoyal}.

For the computation of the self-energy, we also need the Floquet-Wigner transformation of a different type of product given by
$A(t,t') = B(t,t')C(t,t')$. In this case one obtains directly
\begin{align}
 \label{algebraicFloquet}
  A_{nm}(\omega,t)  &=  \sum_{n',m'} \int \frac{d\omega'}{2\pi} ~ B_{n'm'}(\omega',t) \\
 & \qquad \qquad \qquad \times ~ C_{n-n',m-m'}(\omega - \omega',t) \nonumber
\end{align}

\section{Floquet-Boltzmann Equation}
Boltzmann equations are a powerful tool to describe how scattering processes affect the semiclassical dynamics. 
They do not aim at describing quantum-coherent processes at short times, 
but instead focus on the physics at time scales set by slow changes of external parameters and by the scattering time of particles. 
It therefore builds on a clear separation of the time scales for quantum-coherent processes (captured by us within the Floquet approach for periodically driven system) 
and for the semiclassical dynamics which changes occupation functions.

The derivation of the Floquet-Boltzmann equation can be divided into four steps. 
(i) Starting point is the quantum kinetic equation (\ref{QKE}) together with  the calculation of self-energy diagrams, 
which are functionals of the Green's function, see Eq.~(\ref{SigmaK},(\ref{SigmaR}). 
The next goal is to use the separation of time scales. 
Therefore, (ii) one uses the Floquet-Wigner representation, introduced in Sec.~\ref{Sec:FloquetWigner}, for all Green's functions and self energies.
Convolutions, '$\circ$', can be written in terms of a Floquet-Moyal product, (\ref{MoyalFloquet}). 
Using the separation of time scales which implies that terms proportional to $\partial_t$ give small contributions, 
we can (iii) Taylor-expand $e^{-\frac{i}{2} (\partial_t^B \partial_\omega^C-\partial^C_\omega \partial_t^B)}$ to leading order, 
i.e., to linear order on the left-hand and to zeroth order on the right-hand side of the quantum kinetic equation  (\ref{QKE}). 
For problems which are not spatially homogeneous a similar Moyal expansion is also used for the spatial coordinates.
Finally, (iv) the resulting equation is projected onto on-shell processes, e.g., by an integration over frequencies.

The final result of these steps is an equation for the occupation functions $n_{\vec k,\xi}(\vec r,t)$ of the Floquet eigenstates at time $t$. 
Here $\vec k$ is the momentum and $\xi$ includes band- and spin indices. The Floquet states at time $t$ are the eigenstates, $\phi_{t_0,\nu}^n(i)$, 
of the Floquet Hamiltonian (\ref{FloquetHam}) with $\nu=(\vec k,\xi)$ and we have to set $t_0=t$. 
In complete analogy to the treatment of the variable $t_0=t$,
 which we used in Sec.~\ref{sec:Floquet} to deal with the slow time dependence, 
we also allow that the Hamiltonian depends smoothly on the spatial parameter $\vec{r}_0=\vec{r}$. For a lattice model with $n_u$ sites per unit cell,
the  eigenfunctions are calculated in momentum space by diagonalizing
a $n_u (2 N_f+1) \times n_u (2 N_f+1)$ dimensional matrix. 
In the following we will denote the corresponding eigenfunctions in momentum space by $\phi_{t, \vec r,\vec k, \xi}^n(i)$ where $n=-N_f,-N_f+1,...,N_f$ is the Floquet index and $i=1, \dots, n_u$ describes the structure of the Bloch-Floquet wave function within the unit cell. 
Sometimes, we will omit the $t$ and $\vec r$ index to simplify notations and just write $\phi_{\vec k, \xi}^n(i)$.

For the following discussion, we will {\em not} discuss the (main) part of the derivation which is identical for Floquet systems 
and conventional cases, as these are well described in the literature \cite{Rammer1986,Wickles2013} and textbooks \cite{Kamenev2011}. 
Instead, we will only describe those aspects which are different in the Floquet case.

\subsection{Semiclassical dynamics and left-hand side of the Floquet-Boltzmann equation}
The Floquet-Moyal expansion, Eq.~(\ref{MoyalFloquet}), differs from the standard Moyal expansion only by the presence of the extra Floquet indices.
This gives rise to a simple matrix multiplication. 
To be able to describe also situations where the occupation functions are not spatially translational invariant,
 but vary smoothly (on length scales large compared to the lattice spacing), 
one uses a Wigner representation and Moyal expansion for space and momentum degrees of freedom, similar to the one described above for time and frequency variables \cite{Kamenev2011,Rammer1986,Wickles2013}. 
A major difference between the momentum and the frequency dependence is, however, 
that the semiclassical occupation functions depend on the quantum numbers momentum and band index.
but not on frequency and Floquet indices.

A derivation of the conventional collisionless Boltzmann equation based on the quantum kinetic approach,
which (in contrast to previous derivations) 
includes all Berry-phase correction to leading-order, has recently been given by Wickels and Belzig \cite{Wickles2013}. 
One can check (see Appendix~\ref{appendixberry}) that the only difference arising in the Floquet case 
is that all Berry curvatures have to be computed from the Floquet-eigenfunctions introduced in Sec.~\ref{sec:Floquet}, 
\begin{eqnarray}
\Omega^\xi_{\mu \nu}&=&\sum_{i,n} (\partial_\nu \bar{\phi}_{t,\vec k, \xi}^n(i)) (\partial_\mu \phi_{t,\vec k, \xi}^n(i)) \\
&&\qquad \qquad -~(\partial_\mu \bar{\phi}_{t,\vec k, \xi}^n(i)) (\partial_\nu \phi_{t,\vec k, \xi}^n(i)) \nonumber
\end{eqnarray}
where the scalar product involves also a summation over the Floquet index $n$.
Here $\partial_\mu$ and $\partial_\nu$,  $\mu, \nu=(t,r_1,r_2,r_3,p_1,p_2,p_3)$ stands for derivatives in time, space and momentum variables \cite{Bamler2013}.  
These Berry curvatures modify the semiclassical equations of motion in phase space and therefore also the left-hand side of the Boltzmann equation which takes the form
\begin{align}
\label{FBELHS}
(\partial_t + \boldsymbol{\mathcal{F}}_{\vec k,\xi} \nabla_k +  \boldsymbol{v}_{\vec k,\xi} \nabla_r) n_{\vec k,\xi}(\vec r, t)=\left. \partial_t n_{\vec k,\xi}(t)\right|_{\rm coll.}
\end{align}
with \cite{Wickles2013,Sundaram1999}
\begin{align}
 \boldsymbol{v}_{\vec k,\xi} &=  (\mathbbm 1 + \boldsymbol{\Omega}^\xi_{rp})\cdot \nabla_{\vec k} \varepsilon_{\xi} - \boldsymbol{\Omega}^\xi_{pt} +  \boldsymbol{\Omega}^\xi_{pp} \cdot \nabla_{\vec r} \varepsilon_{\xi}  \nonumber \\
 \boldsymbol{\mathcal{F}}_{\vec k, \xi} &= - (\mathbbm 1 + \boldsymbol{\Omega}^\xi_{rp} )\cdot \nabla_r \varepsilon_{\xi} + \boldsymbol{\Omega}^\xi_{rt} + \boldsymbol{\Omega}^\xi_{rr} \cdot \nabla_{\vec k} \varepsilon_{\xi} 
\end{align}
While $\vec \Omega_{rp}$, $\vec \Omega_{rr}$ and $\vec \Omega_{pp}$ are matrices with $r,p=(1,2,3)$, 
$\vec \Omega_{rt}$ and $\vec \Omega_{pt}$ are vectors $(t=1)$.
Note also that $\vec \Omega_{rr}$, $\vec \Omega_{pp}$ can be related to effective magnetic fields, and $\vec \Omega_{rt}$, $\vec \Omega_{pt}$ are referred to effective electric fields.

Since many modern applications of Floquet Hamiltonians \cite{Hauke2012,Struck2013,Aidelsburger2013,Jotzu2014,Aidelsburger2015,Lohse2015} 
have as a goal to realize systems with non-trivial Berry phases, 
it is important to keep track of these effects on the left-hand side of the Boltzmann equation. 
Consider, for example, an interacting Floquet system which heats up as function of time (see Sec.~\ref{sec:Haldane}). 
The (slow) change of occupation functions can trigger a change of the momentum-space Berry curvature $\Omega_{p_1p_2}$, 
a momentum-space `magnetic' field. 
This implies that also corresponding momentum-space `electric' fields $\Omega_{tp_1}$ and $\Omega_{tp_2}$ are generated. 
They can, e.g., induce a macroscopic rotation of the cold-atom system.

\subsection{Scattering and the right-hand side of the Floquet-Boltzmann equation}
To calculate the right-hand side of the Floquet-Boltzmann equation,
we start from the self-energies Eq.~\eqref{SigmaK} and \eqref{SigmaR}.
First, we need an expression for the Green's function. 
Due to the assumed separation of time scales, it is sufficient to evaluate the Green's function using a zeroth-order Floquet-Moyal expansion of the Dyson equation \eqref{Dyson}. 
Furthermore, within our perturbative approach, we do not have to include any self-energy corrections (as the self-energy is already $\propto U^2$). 
Using the Floquet-Wigner representation of both the Hamiltonian and the Green's function, the Dyson equation takes with these approximations exactly the form of Eq.~\eqref{dysonFM}.
This implies that we are allowed to use directly the Green's functions of Eq.~\eqref{GsFloquet} with $n_\nu=n_{\vec k,\xi}(t)$ and $\epsilon_{t_0,\nu} = \epsilon_{t,\vec k, \xi}$.

To evaluate the Floquet-Wigner representation of the self-energies 
Eq.~\eqref{SigmaK},\eqref{SigmaR} and the right-hand side of quantum-kinetic equation \eqref{QKE}, we use the convolution formula Eq.~\eqref{algebraicFloquet} twice. 
For the first line of the formula in Eq.~\eqref{SigmaK}, for example, we obtain after a few steps of simplification a contribution 
of the form
\begin{eqnarray}
 \label{example}
&& \Sigma^{K,1}_{\substack{ij, \sigma\\nm}}(\omega,\vec k) = \\
 &&\phantom{\times}~ \frac{i \pi U^2}{2} \sum_{\eta,\mu,\lambda} \sum_{l,s,u} \int \frac{d\vec p}{(2\pi/a)^d} \, \frac{d\vec q}{(2\pi/a)^d} ~ 
           \Phi^{lsu}_{\begin{subarray}{l}  \eta\mu\lambda,\sigma  \\ \vec k \vec p \vec q \end{subarray}}(n,m,i,j)   \nonumber \\
 &&\times~ \delta(\omega+\epsilon_{\vec q-\vec p,\mu, \bar \sigma}-\epsilon_{\vec k - \vec p,\eta, \bar \sigma}-\epsilon_{\vec q,\lambda, \sigma} + (s - l - u) \Omega ) \nonumber \\
 &&\times ~   (1-2 n_{\vec q - \vec p, \mu, \bar \sigma}) (1-2 n_{\vec k - \vec p, \eta, \bar \sigma})  (1-2 n_{\vec q, \lambda,\sigma}) \nonumber
\end{eqnarray}
where $\bar \sigma=\downarrow$ ($\uparrow$) for $\sigma=\uparrow$ ($\downarrow$), the superscript '$1$' refers to the first line of Eq.~\eqref{SigmaK}, and
\begin{align}
\label{PhiMatrix1}
 &  \Phi^{lsu}_{\begin{subarray}{l}  \eta\mu\lambda,\sigma  \\ \vec k \vec p \vec q,t \end{subarray}}(n,m,i,j) = 
 \sum_{m',m''} 
  \phi_{\vec k - \vec p,\eta, \bar \sigma }^{n+l}(i)
 ~ \bar{\phi}_{\vec k- \vec p,\eta, \bar \sigma}^{m'+l}(j) \nonumber \\[-0pt]
  & \qquad  \times ~ \phi_{\vec q-\vec p,\mu, \bar \sigma}^{ m'+s}(j)
 ~ \bar{\phi}_{\vec q-\vec p,\mu, \bar \sigma}^{ m''+s}(i)
 ~ \phi_{\vec q,\lambda,\sigma}^{ m''+u}(i) 
 ~ \bar{\phi}_{ \vec q, \lambda,\sigma}^{m+u}(j)
\end{align}
where $l,s,u,n,m,m',m''$ are Floquet indices, $\eta, \mu, \lambda$ are band indices and $i,j$ denote sites within the unit cell. 
Here, $(2\pi/a)^d$ is the volume of the Brillouin zone.
We have omitted extra $\vec r$ and $t$ labels which each function obtains to reflect the smooth time and spatial dependencies of the system.

The last remaining step is to evaluate the resulting formula on-shell: 
we multiply the right-hand side of the quantum kinetic equation (\ref{QKE})  by the Floquet-spectral function 
$A_{\vec k,\xi,nm}(\omega)\approx 2\pi {\delta(\omega-\epsilon_{\xi})} \phi^{n}_{\vec k,\xi}(i) \bar{\phi}^{m}_{\vec k,\xi}(j)$ of the state with quantum numbers $\vec k$ and $\xi$, 
integrate over frequencies and trace over Floquet- and space indices.
Note that considering only diagonal, on-shell contributions implies that Boltzmann-type equations cannot describe coherent quantum-oscillations.
The resulting equations are therefore only valid on time scales longer than the decay time of such oscillations. 
This is consistent with our assumptions on the separation of time scales underlying our analysis.  
Furthermore, a quasiparticle has to be well defined, implying that the broadening of the spectral function by scattering is small compared to the energy of the quasiparticles (and therefore also small compared to $\Omega$).

After this last transformation Eq.~(\ref{example}) takes, for example, the form
\begin{eqnarray}\label{exampleC}
 \int&&  \frac{d \omega}{2\pi}\text{Tr} \{ A_{\vec k,\xi}(\omega)  \Sigma^{K,1}(\omega,\vec k) \}\\
=&& \phantom{\times} ~\frac{i \pi U^2}{2} \sum_{\eta,\mu,\lambda} \sum_{l,s,u} \int \frac{d \vec p}{(2\pi/a)^d} \, \frac{d\vec q}{(2\pi/a)^d} ~ 
           \Phi^{lsu}_{\begin{subarray}{l}  \eta\mu\lambda,\sigma  \\ \vec k \vec p \vec q \end{subarray}}(\xi)  \nonumber \\
&& \times~\delta(\epsilon_{\vec k,\xi,\sigma}+\epsilon_{\vec q-\vec p,\mu,\bar \sigma}-\epsilon_{\vec k - \vec p,\eta,\bar \sigma}-\epsilon_{\vec q,\lambda,\sigma} + \Delta_{slu} \Omega ) \nonumber \\[5pt]
&&\times ~   (1-2 n_{\vec q - \vec p, \mu,\sigma'}) (1-2 n_{\vec k - \vec p, \eta,\sigma'})  (1-2 n_{\vec q, \lambda,\sigma}) \nonumber
\end{eqnarray}
with $\Delta_{slu}= s-l-u$, and the transformed matrix element
\begin{equation}
 \Phi^{lsu}_{\begin{subarray}{l}  \eta\mu\lambda,\sigma  \\ \vec k \vec p \vec q\end{subarray}}(\xi)  = 
 \sum_{ij,nm} \bar{\phi}^n_{\vec k,\xi,\sigma}(i) ~
 \Phi^{lsu}_{\begin{subarray}{l}  \eta\mu\lambda,\sigma  \\ \vec k \vec p \vec q \end{subarray}}(n,m,i,j) ~
 \phi^m_{\vec k,\xi,\sigma}(j)
\end{equation}
It is convenient to introduce for each occupation function a separate momentum variable and a $\delta$-function which guarantees momentum conservation (modulo  reciprocal lattice vectors $\vec G_a$).
Performing the entire procedure for all terms of Eq.~\eqref{SigmaK} and likewise for all contributions associated with the second term on the right-hand side of the quantum kinetic equation \eqref{QKE},
one eventually finds an expression for the collision integral 
\begin{align}
\label{CollInt}
&I_{\text{coll}}[n_{\vec k,\xi,\sigma}] = \sum_{\eta,\mu,\lambda} \sum_{\alpha,n} \int \frac{d \vec q_1}{(2\pi/a)^d} \, \frac{d\vec q_2}{(2\pi/a)^d} \, \frac{d \vec q_3}{(2\pi/a)^d} \nonumber\\
& \quad \times  W^{n}_{\begin{subarray}{l}  \xi\mu\eta\lambda,\sigma \\ \vec k \vec q_1 \vec q_2 \vec q_3 \end{subarray}}  ~  (2\pi/a)^d~~ 
\delta(\vec k + \vec q_1 - \vec q_2 - \vec q_3 - \alpha \vec G) \nonumber  \\
& \quad \times ~  \delta(\epsilon_{\vec k,\xi,\sigma}+\epsilon_{\vec q_1,\mu,\bar \sigma}-\epsilon_{\vec q_2,\eta,\sigma}-\epsilon_{\vec q_3,\lambda,\bar \sigma} - n \Omega ) \nonumber \\
& \quad \times ~ \big[ n_{\vec q_2, \eta, \sigma}~ n_{\vec q_3, \lambda, \bar \sigma} ~ (1-n_{\vec k, \xi, \sigma}) ~ (1-n_{\vec q_1, \mu, \bar \sigma}) \nonumber \\
& \quad \qquad\, - ~ n_{\vec k, \xi, \sigma} ~ n_{\vec q_1, \mu, \bar \sigma}~ (1-n_{\vec q_2, \eta, \sigma}) ~(1-n_{\vec q_3, \lambda, \bar \sigma}) \big] 
\end{align}
where we have introduced the integers $\alpha, n \in \mathbbm Z$ to account for Umklapp scattering in momentum- and frequency space, respectively, and 
$ W^{n}_{\begin{subarray}{l}  \xi\mu\eta\lambda,\sigma \\ \vec k \vec q_1 \vec q_2 \vec q_3 \end{subarray}}$
 is the scattering rate for a process involving an energy transfer to the system of $ n \Omega$, $n \in \mathbbm Z$. We obtain
\begin{align}\label{matrixE}
  W^{n}_{\begin{subarray}{l}  \xi\mu\eta\lambda,\sigma \\ \vec k \vec q_1 \vec q_2 \vec q_3 \end{subarray}} =
 2 \pi \,U^2
 \left| V^{n}_{\begin{subarray}{l}  \xi\mu\eta\lambda,\sigma \\ \vec k \vec q_1 \vec q_2 \vec q_3 \end{subarray}} \right|^2 
\end{align}
with the amplitude
\begin{align}\label{amplitude}
  V^{n}_{\begin{subarray}{l}  \xi\mu\eta\lambda,\sigma \\ \vec k \vec q_1 \vec q_2 \vec q_3 \end{subarray}} &= 
 \sum_{i,n_1,n_2,n_3,n_4} \delta_{n-(n_1+n_2-n_3-n_4)}\\
  &\times~  \bar{\phi}^{n_1}_{\vec k,\xi,\sigma}(i) ~ \bar{\phi}^{n_2}_{\vec q_1,\mu,\bar \sigma}(i) 
  ~\phi^{n_3}_{\vec q_2,\eta,\bar \sigma}(i) ~\phi^{n_4}_{\vec q_3,\lambda,\sigma}(i) \nonumber
\end{align}
Note that the Floquet- and momentum indices enter the matrix elements, and therefore the collision integral, in a completely different way:
occupation functions depend on momentum and band indices, but do not depend on the Floquet indices. 
Correspondingly, we sum over Floquet indices in Eq.~(\ref{amplitude}), but not over momentum or band indices. 
We will discuss this important difference again in the concluding section.

The collision integral in Eq.~\eqref{CollInt} forms the right-hand side of the \textit{Floquet-Boltzmann equation}
\begin{align}
\label{FBE}
(\partial_t + \boldsymbol{\mathcal{F}}_{\vec k,\xi} \nabla_{\vec k} +  \boldsymbol{v}_{\vec k,\xi} \nabla_{\vec r}) n_{\vec k,\xi,\sigma}(\vec r, t) = I_{\text{coll}}[n_{\vec k,\xi,\sigma}(\vec r, t)]
\end{align}
where, in general, also the effective forces and velocities depend smoothly on time and space, $\boldsymbol{\mathcal{F}}_{\vec k \xi}=\boldsymbol{\mathcal{F}}_{\vec k,\xi}(\vec r,t)$ 
and $\boldsymbol{v}_{\vec k,\xi}=\boldsymbol{v}_{\vec k,\xi}(\vec r,t)$. 
This dependence can either arise from an explicit $\vec r$ and $t$ dependence of the Hamiltonian or arise from Hartree-Fock corrections to the Hamiltonian, which have to be computed using 
 $n_{\vec k,\xi,\sigma}(\vec r, t)$.

The Floquet-Boltzmann equation (\ref{FBE}) and the formulas for the collision integral (\ref{CollInt}),(\ref{matrixE}),(\ref{amplitude}) are the main results of the first part of the paper.

\section{Haldane model}
\label{sec:Haldane}

\subsection{Model}
In the following we want to apply the Floquet-Boltzmann equation to a concrete example.
In a recent experiment with ultracold atoms in an optical lattice, the Haldane model was realized by means of periodic shaking of the lattice \cite{Jotzu2014}. 
The Haldane model is the prototypical example of a topological insulator: 
Haldane showed that an integer quantum Hall state can be realized {\em without} any external magnetic field on average, 
but just by arranging complex hopping parameters on a hexagonal lattice \cite{Haldane1988}.

\begin{figure}[t]
   \centering
   \includegraphics[width=7cm]{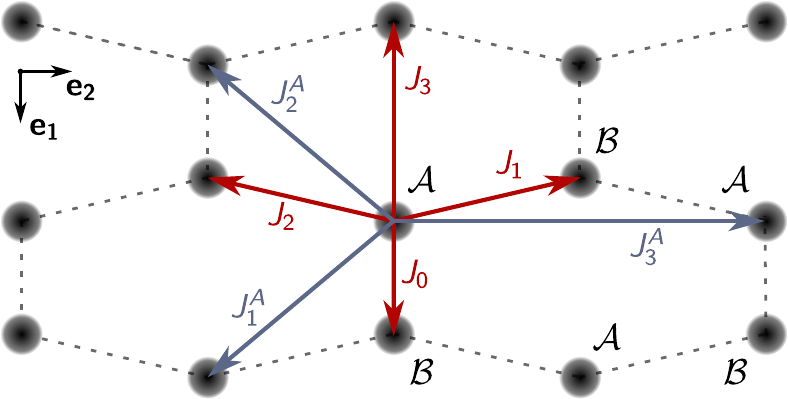}
   \caption{(Color online) Scheme of the hexagonal lattice used to realize the Haldane model \cite{Jotzu2014}.
   For every tunnelling amplitude $t_{j}$ $(t^A_j)$ there is an associated lattice vector $\vec v_j$ $(\vec u_j)$.
   Note that $t^A_j=t^B_j$ and that the phase of complex hopping strengths is defined along the direction of the respective vector in this figure.
   }
 \label{lattice}
\end{figure}

The experiment can be described (see supplementary information of Ref.~\cite{Jotzu2014}) by a (distorted) honeycomb lattices, see Fig.~\ref{lattice}, with two sites per unit cell, which form two
  chequerboard sublattices  $\mathcal A$ and $\mathcal B$. The static Hamiltonian can be described by real nearest-neighbour and next-nearest-neighbour hopping amplitudes
\begin{align}
 H=& \sum_{\vec u \in \mathcal A,\sigma}  \Big[  \frac{\Delta_{AB}}{2} 
 ( a_{\vec u\sigma}^{\dagger} a_{\vec u\sigma}^{\phantom{\dagger}} - b_{\vec u+\vec v_0,\sigma}^{\dagger} b_{\vec u+\vec v_0,\sigma}^{\phantom{\dagger}} )
 \\ +& \sum_{j,\sigma} (J_j b_{\vec u + \vec v_j,\sigma}^{\dagger} a_{\vec u\sigma}^{\phantom{\dagger}} + \text{h.c.} ) \nonumber \\
 +& \sum_{j',\sigma} ( J_{j'}^{A} a_{\vec u + \vec u_{j'},\sigma}^{\dagger} a_{\vec u\sigma} + J_{j'}^{B} b_{\vec u + \vec v_0 + \vec u_{j'},\sigma}^{\dagger} b_{\vec u + \vec v_0,\sigma} + \text{h.c.}) \Big] 
 \nonumber \label{H}
\end{align}
with $\sigma$ being a spin index, and vectors $\vec u_j$ connecting points on the same sub-lattice and vectors $\vec v_j$ that connect points on different sub-lattices.

A periodic shaking of the lattice leads to an acceleration of all atoms. 
In the frame of reference comoving with the lattice, an acceleration can be viewed by a force  $\vec F(t)$, which is constant in space. 
Due to the periodic shaking, the force is periodic in time, $\vec F(t) = \vec F (t+\ttt)$.  
Within one period $\ttt$, $\vec F(t)$ rotates in the plane of the lattice on an ellipse. 
The force can either be implemented in the Hamiltonian by a potential or, more conveniently, by a vector potential 
$\vec A(t) = - \frac{ 2 \hbar K_0 }{ \lambda}(\sin (\Omega t) \vec e_1 + \sin (\Omega t - \varphi) \vec e_2 )$ 
with $\partial_t \vec A(t)=\vec F(t)$. 
Here, $\lambda$ is the wave-length of the laser used to create the optical lattice, 
$K_0 = 0.7778$ parametrizes the strength of shaking in the two perpendicular directions  $\vec e_1$ and $\vec e_2$, 
and $\Omega$ is the oscillation frequency used in the experiment.  The parametrization is chosen such that $\lambda$ will cancel in the final result.

The vector potential modifies the hopping amplitudes by complex phases.
Using Eq.~(\ref{H}), this leads to a description of the system  with complex hopping amplitudes depending periodically on time, 
$J_j(t+\ttt)=J_j(t)$, $J_j^{A/B}(t+\ttt)=J_j^{A/B}(t)$ where
\begin{align}
J_j &= e^{i z_{j} \sin (\Omega t+ \phi_{j})} J_{j} \\
J^A_j &= e^{i z^A_{j} \sin (\Omega t+ \phi_{j}^A)} J^A_{j} \nonumber \\
J^B_j &= J^A_j 
\nonumber
\end{align}
with 
\begin{align}
z_{j}^{(A)} &= \frac{2 K_0}{\lambda} \rho_{j}^{(A)}\\
\rho_{j} e^{i \phi_{j}} &= \vec v_{j} \cdot \vec e_1 + \vec v_j \cdot \vec e_2 ~ e^{- i \varphi} \nonumber \\
\rho_{j}^A e^{i \phi_{j}^A} &= \vec u_{j} \cdot \vec e_1 + \vec u_j \cdot \vec e_2 ~ e^{- i \varphi} \nonumber
\end{align}
and $\rho_{j} \geq 0$. 
We have implemented the equations for two sets of parameters. 
In appendix~\ref{alternativepara}, we show the results for the microscopic Hamiltonian studied in the main text of Ref.~\cite{Jotzu2014},
which is anisotropic and includes both nearest-neighbor and next-nearest neighbor interactions. 
All qualitative features are, however, unchanged compared to a simpler set of parameters investigated in the following. 
To mimic the situation discussed in the supplementary material of Ref.~\cite{Jotzu2014} (see also Ref.~\cite{Uehlinger2013}), where a spinfull, interacting system has been studied, 
we set  $J_0= J_1=J_2= J= - 2 \pi\cdot172$\,Hz, and use that approximately $J_0 = J_{i}^A =\Delta_{AB}=0$ and set $\phi=\pi/2$, thus describing a situation where the ground state is in the (quantum-Hall) topological phase (note that in the experiment also a hopping coupling adjacent honeycomb layers was present, which is, however, ignored here). 
Furthermore, we use \cite{Jotzu2014}  $\vec v_0= \lambda (0.438, 0)$, $\vec v_1= \lambda (-0.062, 0.5)$ and $\vec v_2=\lambda (-0.062, -0.5)$. We only consider the case of a translationally invariant system at half filling.

%

To analyze theoretically the effective Hamiltonian arising from the shaking of the lattice, the authors of \cite{Jotzu2014} used the so-called Magnus expansion,
equivalent to a determination of the Floquet eigentstates to first order perturbation theory in $1/\Omega$. Energy non-conserving processes can, however, not be treated with a simple Magnus expansion.

We will now consider the effects of interactions by adding to the Hamiltonian the term
\begin{eqnarray}
H_{\rm int}&= U \sum_i n_{i \uparrow} n_{i \downarrow}
\end{eqnarray}
with $n_{i\sigma}=  a_{\vec u_i\sigma}^{\dagger} a_{\vec u_i\sigma}^{\phantom{\dagger}}$ ($n_{i\sigma}=  b_{\vec u_i+\vec v_0,\sigma}^{\dagger} b_{\vec u_i+\vec v_0,\sigma}^{\phantom{\dagger}}$ ) 
on the $\mathcal A$ ($\mathcal B$) sublattice, respectively. 
While in Ref.~\cite{Jotzu2014} mainly the non-interacting spinless case was considered, 
the authors also briefly studied the spinfull limit where the interplay of interactions and periodic modulations is expected to heat up the system. 

The Hartree correction turn our to give only tiny contributions. 
It leads to a periodically oscillating term $\Delta_{AB}(t)\approx  c\, U \cos(\Omega t)$ where  $c\lesssim 0.2$ depends on the occupation function of all states. 
The Hartree contribution therefore remains  small compared to all other terms for values of $U \lesssim J$ where our perturbative formulas can be applied. 
The single-particle gap, for example, changes only by $0.2 \%$ for $U=J$ and $c=0.2$. 
We have therefore neglected the Hartree correction which considerably simplifies the numerics as all scattering matrix elements have  to be computed only once.
Furthermore, this approximation implies that the dependence of $U$ can be absorbed into a redefinition of the time, see below.

To determine the Floquet eigenstates and energies for the Floquet-Boltzmann equation it is therefore sufficient to diagonalize the non-interacting Floquet Hamiltonian given by Eq.~\eqref{FloquetHam}, 
where the Fourier components $h^0_n$ (for each spin species) can be written as $2\times2$ matrices (formulated in momentum space) 
\begin{eqnarray}
 h_n^0=h_{n,i}^0 \mathbbm 1 + h_{n,x}^0 \sigma_x+ h_{n,y}^0 \sigma_y
\end{eqnarray}
where $\sigma_i$ are Pauli matrices and
\begin{align}
 h^0_{n,i} &= \sum_{j'} J_n(z_{j'}) \, J_{j'} e^{i n \phi_{j'}} (e^{i \vec q \cdot \vec u_{j'}} + (-1)^n e^{-i \vec q \cdot \vec u_{j'}})  \nonumber \\
 h^0_{n,x} &= \frac{1}{2}\sum_{j} J_n(z_{j})\, J_{j} e^{i n \phi_{j}} (e^{i \vec q \cdot \vec v_{j}} + (-1)^n e^{-i \vec q \cdot \vec v_{j}})   \\
 h^0_{n,y} &= \frac{1}{2i}\sum_{j} J_n(z_{j}) \,J_{j} e^{i n \phi_{j}} (e^{i \vec q \cdot \vec v_{j}} - (-1)^n e^{-i \vec q \cdot \vec v_{j}}) \nonumber
\end{align}
Here, $J_n(\cdot)$ describes the Bessel function of the first kind to $n$th order. 
Note that the formula for $h^0_{n,i}$ corrects a typo in the 
supplement of \cite{Jotzu2014}.
Numerical diagonalization of the Floquet Hamiltonian yields then the Floquet states $\phi^n_{\vec k, \xi}$ and energies $\epsilon_{\vec k, \xi}$.

\begin{figure}[t]
   \centering
   \includegraphics{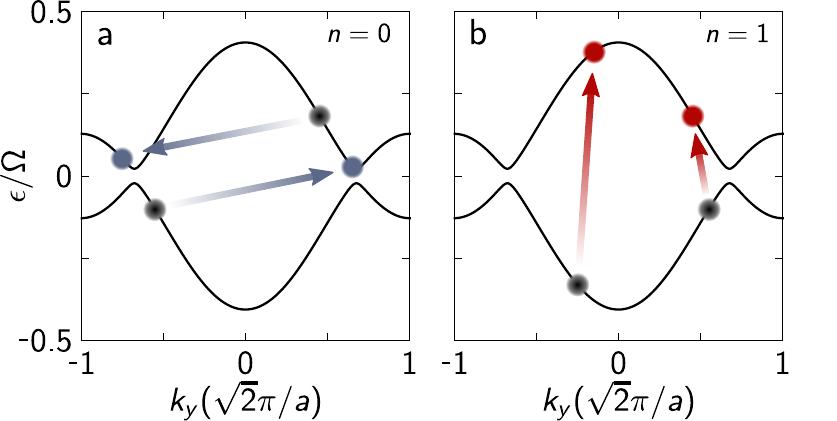}
   \caption{(Color online). Two-particle collision processes for (a) the energy conserving case $(n=0)$ and (b) the energy violating case $(n=1)$.
   The energy bands of the system described in the main text ($\Omega=6.3 |J|=2 \pi\cdot 1080$\,Hz) are  shown along the diagonal of the quadratic Brillouin zone ($k_x=0$ within our conventions).
   Scattering is indicated by arrows from initial states (black) into final states (blue/red).
   }
 \label{bandprocesses}
\end{figure}

\subsection{Quasi-equilibrium and heating rate}

We consider a translationally invariant situation without external forces.
Hence, the Floquet-Boltzmann equation in Eq.~\eqref{FBE} reduces to the simplified form 
\begin{equation}
\partial_t n_{\vec k,\xi,\sigma}(t) = I_{\text{coll}}[n_{\vec k,\xi,\sigma}]
\end{equation} 
with $n_{\vec k,\xi,\uparrow}(t)=n_{\vec k,\xi,\downarrow}(t)=n_{\vec k,\xi}(t)$.
The Floquet states and quasi energies appearing in the collision integral are time-independent.


To study the heating of the system, we consider the change of the (Floquet-) energy par lattice site defined by
\begin{equation}
E(t)=\frac{1}{2}\int \frac{d\vec k}{(2\pi/a)^2} \sum_{\xi,\sigma} \epsilon_{\vec k,\xi} ~ n_{\vec k, \xi,\sigma}(t) 
\end{equation}
where the factor $1/2$ arises as there are two lattice sites per unit cell.
The \textit{heating rate}, $\gamma = dE/ dt$, is therefore given by
\begin{eqnarray}
\label{heating1}
\gamma(t)=\frac{dE(t)}{dt} = \frac{1}{2}\int \frac{d\vec k}{(2\pi/a)^2} \sum_{\xi,\sigma} \epsilon_{\vec k,\xi} ~ \dot{n}_{\vec k, \xi,\sigma}(t)
\end{eqnarray}
Using the right-hand side of the Floquet-Boltzmann equation \eqref{FBE} and the symmetry properties of the scattering rates, the heating rate can be written as
\begin{align}
\label{heating2}
\gamma(t) & = \sum_{\xi,\eta,\mu,\lambda} \sum_{n,\alpha}  \int \frac{d\vec k}{(2\pi/a)^2}\, \frac{d \vec q_1}{(2\pi/a)^2} \, \frac{d\vec q_2}{(2\pi/a)^2} \, \frac{d\vec q_3}{(2\pi/a)^2} \nonumber \\
& \times n \Omega\, W^{n}_{\begin{subarray}{l}  \xi\mu\eta\lambda,\sigma \\ \vec k \vec q_1 \vec q_2 \vec q_3 \end{subarray}}\,  \delta(\epsilon_{\vec k,\xi}+\epsilon_{\vec q_1,\mu}-\epsilon_{\vec q_2,\eta}-\epsilon_{\vec q_3,\lambda} - n \Omega ) \nonumber \\
&\times \frac{1}{2} ~ (2\pi/a)^2~ \mathcal 
\delta(\vec k + \vec q_1 - \vec q_2 - \vec q_3 - \alpha \vec G) \nonumber \\
& \times ~ n_{\vec q_2, \eta}(t) ~ n_{\vec q_3, \lambda}(t) ~ (1-n_{\vec k, \xi}(t)) ~ (1-n_{\vec q_1, \mu}(t)) 
\end{align}
The energy changes in quanta of $n \Omega$, determined by the Floquet matrix elements 
$W^{n}$ and the occupation functions.

\begin{figure}[t]
   \centering
   \includegraphics{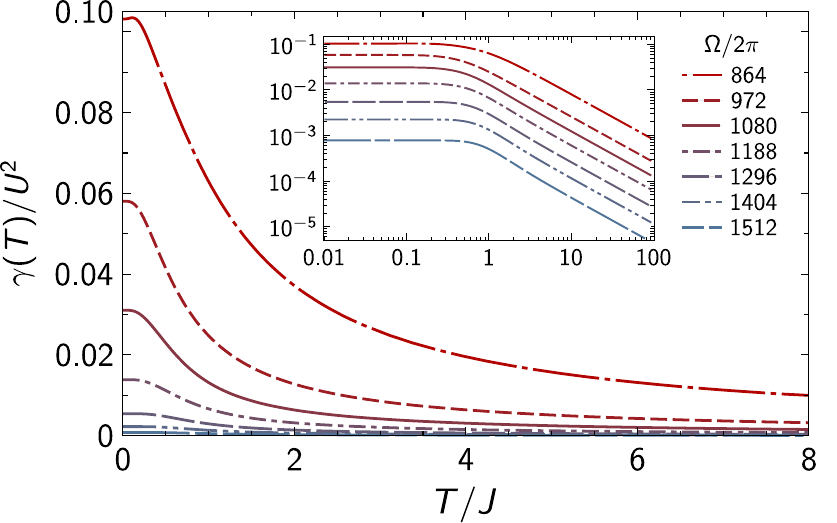}
   \caption{(Color online) Heating rate per lattice site (in units of $1/U^2$) plotted against dimensionless temperature $T/J$, 
   where $J$ is the hopping amplitude of the isotropic Hubbard model at hand.
   Different curves describe different driving frequencies $\Omega$, ranging from $5 |J|=2 \pi\cdot 864$\,Hz to $8.8 |J|=2 \pi\cdot 1512$\,Hz, see legend.
   The inset shows a double logarithmic plot.}
 \label{g_vs_T}
\end{figure}

The occupation functions $n_{\vec q_2, \eta}(t)$ can be determined from the solution of the Floquet-Boltzmann equation. 
This is in general a formidable task due to the high-dimensional integrals occurring in the collision integral~(\ref{CollInt}). 
The problem can, however, be simplified dramatically in situations where the scattering rate, $1/\tau_{\rm con}$, 
of energy-conserving scattering processes ($n=0$) dominates over the scattering rate, $1/\tau_{\rm vio}$, for processes which violate energy conservation
\begin{equation}\label{tau}
\frac{1}{\tau_{\rm con}} \gg \frac{1}{\tau_{\rm vio}}
\end{equation}
In Fig.~\ref{bandprocesses} we depict two typical processes associated with both time scales.
For the parameters studied experimentally in Ref.~\cite{Jotzu2014} these rates differ by much more than an order of magnitude as discussed below.

Energy-conservating processes lead to equilibration. 
This implies that under the condition of Eq.~(\ref{tau}) after a few energy-conserving scattering events the occupation functions are well approximated by thermal ones
\begin{equation}
\label{nequ}
n_{\vec k, \xi}(t) \approx n^{0}_{\vec k, \xi}(T(t)) =\left(\exp\!\left[\frac{\epsilon_{\vec k \xi}-\mu(T(t))}{T(t)}\right]+1\right)^{-1}
\end{equation} 
where $\mu(T)$ is generally determined from the condition that the total number of particles remains constant.
Note also that we set $k_B=\hbar=1$.

To leading order in  $\tau_{\rm vio}/\tau_{\rm con}$ we can therefore replace all occupation functions both in Eq.~\eqref{heating1} and Eq.~\eqref{heating2} by Fermi functions, $n(t) \to n^0(T(t))$.
We therefore obtain
\begin{eqnarray}
\label{dgl}
\frac{d T(t)}{d t} \frac{d E}{d T} \approx \gamma(T(t))
\end{eqnarray}
which can directly be solved by 
\begin{equation}
\label{t}
 t = \int_{T_i}^{T(t)} dT' ~ \frac{c(T')}{\gamma(T')}
\end{equation}
where $c(T)=d E/d T$ is the specific heat.
Hence, under condition (\ref{tau}) we do not have to solve the 
complicated coupled integro-differential equations (\ref{FBE}) for $n_{\vec k, \xi}(t)$. 
Instead, it is sufficient to calculate $\gamma$ and $E$ as function of temperature and to solve a simple one-parameter differential equation (\ref{dgl}). 
Note that the high-dimensional integral \eqref{heating2} can still be numerically demanding.

%
%

To determine $\gamma(T)$, one first has to calculate the matrix elements. For the experimental parameters the Floquet eigenstates are well localized.
For driving frequencies of $\Omega \gtrsim 3 |J|\approx 2 \pi \cdot 500\, \rm Hz$ we have checked  that it is sufficient to keep track of a few Floquet modes only, e.g., 
$N_F = 3$. Note that $W^{n}$ with $|n|\ge 2$ cannot contribute to Eq. (\ref{heating2}) as $|\epsilon_{\vec k \xi}|\le \Omega/2$.
To calculate the integrals in (\ref{heating2}) taking into account energy conservation, 
we discretize not only momentum by a $20 \times 20$ mesh but we also discretize the energies by rounding them to multiples of $\Delta E=2 \pi \cdot 12\rm Hz\approx 0.07 |J|$. 
We have checked that for these parameters discretization errors are negligible.

\begin{figure}[t]
   \centering
   \includegraphics{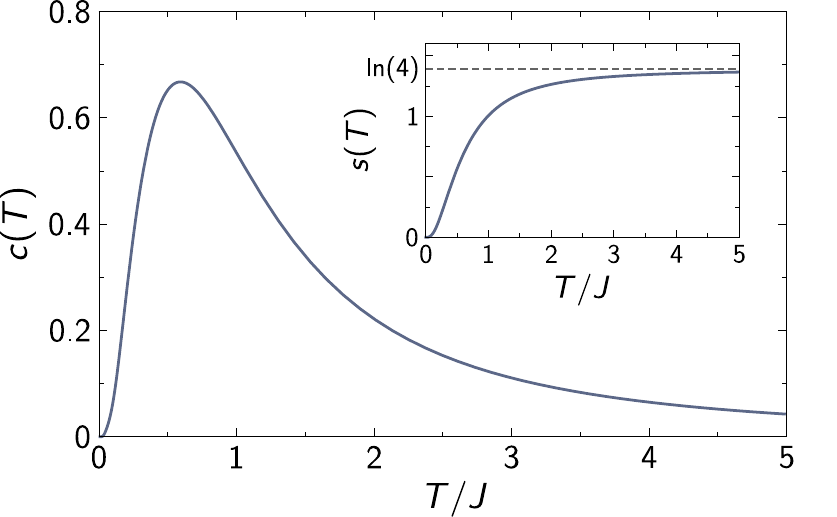}
   \caption{(Color online) Specific heat per lattice site, $c(T)$, plotted as function of dimensionless temperature $T/J$ for a driving frequency $\Omega =6.3 |J|=2 \pi \cdot 1080 \rm{Hz}$.
   The inset shows the corresponding entropy per lattice site $(k_B = 1)$ as function of $T/J$. $c(T)$ depends only very weakly on $\Omega$ in the considered parameter regime.  }
 \label{c_vs_T}
\end{figure}

Fig.~\ref{g_vs_T} shows the heating rate $\gamma(T)$  for different driving frequencies $\Omega$.
One observes that all heating rates show the qualitative same behaviour:
$\gamma$ starts out at some maximal value at $T=0$ and then approaches $\gamma=0$ for $T\to\infty$.
To analyze the limit $T \to \infty$, it is useful to realize that in the situation where (\ref{nequ}) holds, 
a detailed balance condition relates the rate $\gamma^+$, which is defined by the terms in $\gamma$ proportional to $W^{+1}$, to $\gamma^-$, the terms proportional to $W^{-1}$
\begin{eqnarray}
\frac{\gamma^+}{\gamma^-}=e^{\Omega/T}
\end{eqnarray}
This follows from the identity for Fermi functions $
 n^0_\eta n^0_\lambda (1-n^0_\xi)(1-n^0_\mu)=n^0_\xi n^0_\mu (1-n^0_\eta)(1-n^0_\lambda) e^{ \Omega \beta }$ for $\epsilon_{\xi}+\epsilon_{\mu}-\epsilon_{\eta}-\epsilon_{\lambda} = \Omega$.
As $\gamma=\gamma^+ - \gamma^-$, we obtain
\begin{eqnarray}
\gamma(T) \propto \frac{1}{T}\label{gammaA}
\end{eqnarray}
for $T \gg \Omega$.

For increasing driving  frequency, the heating rate drops rapidly, see Fig.~\ref{g_vs_T}. 
This has two reasons: first, the matrix elements $W^{\pm 1}$ drop for increasing $\Omega$ with  $W^{\pm 1} \propto \frac{1}{\Omega^2}$ for $\Omega \to \infty$. 
Second, the phase space for two-particle scattering with energy transfer $\pm \Omega$ vanishes due to the restrictions on energy conservation for $\Omega>2 D(\Omega)$.
Here, $D(\Omega)$ is the total bandwidth, $D(\Omega)=\max_{\vec k,\xi} \epsilon_{\vec k \xi} - \min_{\vec k,\xi} \epsilon_{\vec k \xi}$. 
By expanding around the band minima and maxima, we obtain 
\begin{eqnarray}
\gamma \propto (\Omega_{\rm max}-\Omega)^3
\end{eqnarray}
 for $\Omega \to \Omega_{\rm max}$, $\Omega<\Omega_{\rm max}$ with $2 D(\Omega_{\rm max})=\Omega_{\rm max}$. 
 As for $\Omega>\Omega_{\rm max}$, two particle scattering cannot contribute to heating, one has to consider higher-order scattering events. 
 Within our model, we obtain $\Omega_{\rm max}=2\pi\cdot 1778.5\,\rm Hz=10.34\,|J|$.

\begin{figure}[t]
   \centering
   \includegraphics{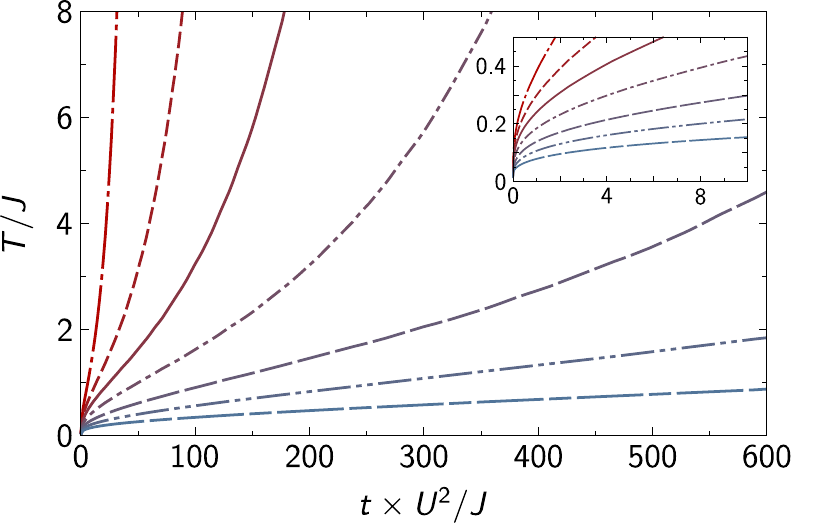}
   \caption{(Color online) Dimensionless temperature, $T/J$, of the interacting system as a function of time for different driving frequencies $\Omega$ ($\Omega=2 \pi \times 864, \dots, 1512$\,Hz), 
   see legend of Fig.~\ref{g_vs_T}.  
   The inset clarifies that the curves undergo two distinct regimes before rising with exponential speed. 
   }
 \label{T_vs_t}
\end{figure}

The specific heat per lattice site is determined from
\begin{align}
\label{heatcapacity}
c(T)  & = \frac{d E(T)}{d T} =  \int \frac{d\vec k}{(2\pi/a)^2} \sum_{\xi}  \epsilon_{\vec k,\xi} \,\frac{d}{d T} n^0_{\vec k,\xi}(T) \nonumber
\end{align}
Fig.~\ref{c_vs_T} shows the specific heat as a function of temperature. 
For $T \to 0$ the specific heat is exponentially suppressed due to the band gap $\Delta_G$ with  $\Delta_G\approx 0.3 \,J$ for $\Omega=2 \pi \cdot 1080\,\rm Hz$ 
(for all $\Omega$ considered by us the system remains in the gapped topological state). 
For $\Delta_G < T < J$, the system is approximately described by a Dirac equation and $c(T)$ grows with $T^2$.
For $T\gg J$, in contrast, one obtains $c(T) \propto 1/T^2$.

Using either Eq.~\eqref{dgl} or Eq.~\eqref{t}, we can directly compute the evolution of temperature as function of time (assuming $T_i=0$ as the initial temperature). 
This is shown in Fig.~\ref{T_vs_t} for different values of the driving frequency $\Omega$.

Due to the exponential suppression of $c(T)$ for $T\to 0$, initially $T(t)$ rises very rapidly proportional to $1/\log[1/t]$ for $T(t)$ small compared to the band-gap $\Delta_G$, followed by $T(t) \propto t^{1/3}$ for $\Delta_G < T < J$ (see inset of  Fig.~\ref{T_vs_t}). 
For larger temperatures, there is an intermediate regime with an approximately linear rise of $T$. 
Finally, for $T\gg \Omega$, we obtain $d T/d t\propto T$ from $\gamma \propto 1/T$, $c(T)\propto 1/T^2$ and Eq.~\eqref{dgl},  and therefore an exponential rise of the temperature
\begin{equation}\label{Tas}
 T(t)\propto e^{c t} \quad \text{for } T \gg \Omega
\end{equation}
as confirmed numerically.

Due to the strong dependence of $\gamma$ on the driving frequency $\Omega$, also $T(t)$ depends strongly on $\Omega$. 
For $\Omega \to \Omega_{\rm max}$, for example, the prefactor $c$ in Eq. (\ref{Tas}) descreases rapidly, $c\propto (\Omega_{\rm max}-\Omega)^3$.

\begin{figure}[t]
   \centering
   \includegraphics{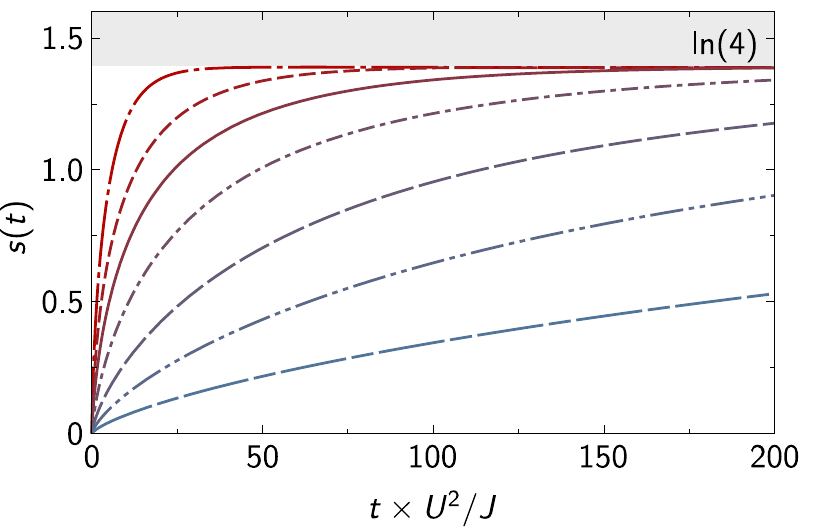}
   \caption{(Color online) Entropy per lattice site (per particle) as a function of time for different driving frequencies $\Omega$ ($\Omega=2 \pi \times 864, \dots, 1512$\,Hz), 
   see legend of Fig.~\ref{g_vs_T}. 
   The solid curve ($\Omega  =6.3 |J|= 2 \pi \cdot 1080$\,Hz) is the lowest frequency used in the experiment \cite{Jotzu2014}.
}
 \label{s_vs_t}
\end{figure}

In cold atom systems, it is not easy to determine the temperature of the system. 
An observable, which is sometimes better accessible is the entropy per particle (also determined in the extended data section of Ref.~\cite{Jotzu2014}). 
The reason is that the entropy remains  approximately constant when all optical lattices and interactions are slowly switched off. 
We therefore show in Fig.~\ref{s_vs_t} also the entropy as a function of time. 
Due to the exponential rise of $T$ with time, Eq. (\ref{Tas}), the entropy approaches its $T\to \infty$ limit with exponential speed, 
$S \approx \ln[4]-c_0/T^2\approx  \ln[4]- c_1 e^{-2 c t}$, for $t\to \infty$. 

\begin{figure}[t]
   \centering
   \includegraphics{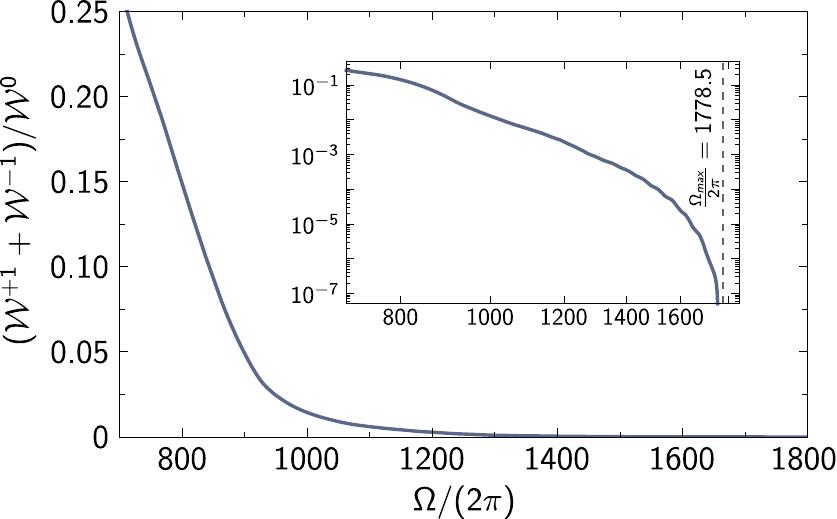} 
\caption{The ratio of scattering times, Eq.~(\ref{ratio2}), for energy non-conserving $(W^{\pm1})$ and energy conserving processes $(W^{0})$ plotted as a function of the driving frequency $\Omega$.
For large frequencies, $\Omega > 6 |J|\approx 2 \pi\cdot 1000$\,Hz energy conserving scattering dominates by several orders of magnitude. 
 Inset: double-logarithmic plot.  }
 \label{ratios}
\end{figure}

Finally, we have to check whether in the experimental system the condition on the scattering rates (\ref{tau}) is fulfilled. 
Taking into account the high-temperatures of the experimental system, 
we will check the condition by comparing the ratio of the scattering rates averaged over all bands and momenta in the limit $T \to \infty$. 
We define 
\begin{align}
\label{ratio1}
\mathcal W^{n} & = \sum_{\xi,\eta,\mu,\lambda} \sum_{\alpha}  \int \frac{d\vec k}{(2\pi/a)^2}\, \frac{d \vec q_1}{(2\pi/a)^2} \, \frac{d\vec q_2}{(2\pi/a)^2} \, \frac{d\vec q_3}{(2\pi/a)^2} \nonumber \\
& \times W^{n}_{\begin{subarray}{l}  \xi\mu\eta\lambda,\sigma \\ \vec k \vec q_1 \vec q_2 \vec q_3 \end{subarray}}\,  \delta(\epsilon_{\vec k,\xi}+\epsilon_{\vec q_1,\mu}-\epsilon_{\vec q_2,\eta}-\epsilon_{\vec q_3,\lambda} - n \Omega ) \nonumber \\
&\times  \mathcal \delta(\vec k + \vec q_1 - \vec q_2 - \vec q_3 - \alpha \vec G) 
\end{align}
and estimate
\begin{align}
\label{ratio2}
\frac{\tau_{\rm con}}{\tau_{\rm vio}} \approx \frac{\mathcal W^{+1}+\mathcal W^{-1} }{\mathcal W^{0}}
\end{align}
In Fig.~\ref{ratios} we show numerical values for the ratio \eqref{ratio2} as a function of $\Omega$. 
For the (smallest) experimental values, $\Omega \approx 2 \pi\cdot 1000$\,Hz, 
the energy conserving processes already dominate by almost two orders of magnitude, which justifies the approximation of Eq.~(\ref{nequ}) with high precision. 
As discussed above, all two-particle processes violating energy conservation die out with $(\Omega_{\rm max}-\Omega)^3$  for $\Omega \to \Omega_{\rm max}$, see inset of \eqref{ratio2}.

\section{Discussion and Conclusions}

In this paper, we have derived the Floquet-Boltzmann equation for periodically driven Fermi systems starting from the Keldysh-dynamics of Green functions.

It is instructive to compare the effect of the breaking of translational symmetry in time by oscillations periodic in time 
with the effect of the breaking of translational symmetry in space by a potential periodic in space. Many effects are similar: 
in the first case  energy is conserved only modulo $\hbar \Omega$, while in the latter case momentum is conserved only modulo reciprocal lattice vectors, 
$\hbar \vec G$. This leads to a heating of the system to infinite temperature and to the decay of any macroscopic momenta, respectively. 
In both cases the system relaxes to a state with maximal entropy allowed by the remaining conservation laws. 
An important difference is, however, that in the case of a periodic potential, one obtains a nominally infinite number of electronic bands, 
each of which has to be described by an extra quantum number and a corresponding semiclassical occupation function.
In the Floquet case, in contrast, we used the same quantum numbers as in the energy conserving case and did not introduce any new occupation function. 
This can be traced back to the very different role taken by space  and time in single-particle quantum mechanics: 
while the first one is promoted to an operator, this is not the case for time.

The perturbative Floquet-Boltzmann equation derived by us can also be extended to the limit of strong interactions in situations where the number of excitations 
(e.g., doublon and holon excitations in a bosonic Mott insulator) remains small. 
In this case, the transition rates on the right-side of the Boltzmann equation have to be computed from the solution of the 2-particle scattering problem in the presence of periodic perturbation.

The absence of energy conservation, ultimately heats up closed quantum systems to infinite temperature. 
Our calculations have shown that this effect can be quite strong for experimentally relevant parameters \cite{Jotzu2014}. 
Consider, for example, in Fig.~\ref{s_vs_t} the case $\Omega=6.3|J|=2 \pi \cdot 1080$\,Hz for moderate interactions, $U=|J|$.
In this case, the entropy per site rises from $0.5\,k_B \ln[4]$ to $0.75\,k_B \ln[4]$ within $\Delta t \approx 18/J \approx 20$\,ms 
which is short compared to the loading times used in the experiment of Ref.~\cite{Jotzu2014}. 
This clearly shows the importance of the heating effects.
A direct quantitative comparison to the heating rates observed in the experiment ~\cite{Jotzu2014} is, however, not possible. 
In Ref.~\cite{Jotzu2014} rather large values of $U=10 J$ (and $U=20 J$), close to (or in) the Mott insulating phase, were investigated, which are far beyond the applicability of our approach.
Also, a three-dimensional coupling was finite in this experiment. 
For a quantitative description of the heating rate in the experimental setup it would also be necessary to treat the trapping potential which leads to inhomogeneous heating and heat transport through the trap. 
The entropy increase of only a few $k_B/s$ reported in Ref.~\cite{Jotzu2014} appear to be rather small compared to the values estimated by us for small initial entropies. 
We believe that this can only be explained by the fact that the initial entropy per site in the lattice was rather high. 
Indeed, in Ref.~\cite{Uehlinger2013}, which describes a similar setup without shaking, 
entropies {\em per particle} of $1.5\,k_B\approx k_B \ln[4]$ and $2.5\,k_B \approx 1.8 \ln[4]$ have been reported before and after loading the system into the trap, respectively.

As we have shown, the heating rate can efficiently by controlled by moderate changes of the driving frequency. 
For $\Omega$ larger than twice the total bandwidth heating by two-particle collisions is completely absent and only higher-order processes (not included in our analysis) can take place which are expected to be strongly suppressed by the Pauli principle. 
Increasing the driving frequency implies, however, that all effects of the periodic modulation are also suppressed. For the model considered by us, the gap of the topological insulator scales with $1/\Omega$ for large $\Omega$. 
Increasing the driving frequency from $6.3\,|J|\approx 2 \pi \cdot 1080$\,Hz to $\Omega_{\rm max}\approx 10.3 \,|J|\approx 2 \pi \cdot 1800$\,Hz, 
where 2-particle processes are completely suppressed, reduces the gap from $0.3\,J$ to $0.15\,J$.

For the  design of interacting Floquet systems it will be important to control the heating processes. 
Here we hope that the Floquet-Boltzmann equation and variants thereof can be a useful tool.

\begin{acknowledgments}
We thank E. Berg, T. Esslinger  and R. Bamler for useful discussions.
 This work is financially supported by the Deutsche Telekom Stiftung and the Bonn-Cologne Graduate School of Physics and Astronomy (M.G.). 
\end{acknowledgments}

%
%
%
%
%
%
%
%
%
%
%
%

\appendix
\section{Floquet-Moyal expansion}
\label{appendixMoyal}
In the following we sketch the explicit derivation of the Floquet-Moyal expansion.
Starting point is the time-convolution of two two-point functions, given by
\begin{align}
 A(t,t') 
 &=   \int dt_1 ~ B(t,t_1) C(t_1,t') \label{Moyal1} \\
 &= \sum_{n,m} \int \frac{d\omega}{2\pi} ~ e^{-i(\omega + n\Omega)t} e^{i(\omega + m\Omega)t'}~ A_{nm}(\omega,\bar{t}) \label{Moyal2}
\end{align}
In order to find the explicit form of $A_{nm}(\omega,\bar{t})$ one inserts the inverse Floquet-Wigner expressions (see. Eq.~\eqref{InversefloquetWigner}) for $B$ and $C$ into Eq.~\eqref{Moyal1}.
The formula then reads
\begin{align}
 \begin{split}
  A(t,t') =  \sum_{\substack{n',m'\\n'',m''}} &\int \frac{d\omega'}{2\pi} \, \frac{d\omega''}{2\pi} ~ e^{-i(\omega' + n'\Omega)t}~e^{i(\omega'' + m''\Omega)t'} \\[-10pt]
   \times ~ & \int dt_1 ~ e^{i(\omega' + m'\Omega)t_1} ~ e^{-i(\omega'' + n''\Omega)t_1} \\
  \times ~ & B_{n'm'}(\omega',{\scriptstyle\frac{t+t_1}{2}})~ C_{n''m''}(\omega'',{\scriptstyle \frac{t_1+t'}{2}})
 \end{split}
\end{align}
The first technical challenge enters now via the convolution in the time-argument $t_1$.
To make progress here, one expands the functions $B$ and $C$ around $t_1 = t'$ and $t_1 = t$, respectively.
This leads to the following expression
\begin{align}
\begin{split}
 A(t,t') &=  \int \frac{d\omega'}{2\pi} \, \frac{d\omega''}{2\pi} \sum_{\substack{n',m',l\\n'',m'',l'}}  \sum_{k=0}^{l} \sum_{k'=0}^{l'} 
 \frac{1}{l! l'!}\frac{1}{2^{l} 2^{l'}} \binom{l}{k} \binom{l'}{k'}\\
 &\times ~  (-1)^{k} (-1)^{k'} t^{k} t'^{k'} e^{-i(\omega' + n'\Omega)t} e^{i(\omega'' + m''\Omega)t'} \\ 
 &\times ~ \int dt_1 e^{i(\omega'-\omega'' + (m'-n'')\Omega)t_1}~ t_1^{l+l'-k-k'} \\
 &\times ~  B_{n'm'}^{(l,0)}(\omega',\bar{t})~ C_{n''m''}^{(l',0)}(\omega'',\bar{t})
\end{split}
\end{align}
where the identity $(t_1-t)^{l} = \sum_{k=0}^l \binom{l}{k} (-1)^{k} t_1^{l-k} t^{k}$ has been used, 
and $B^{(l,0)}$ describes the $l$th ($0$th) derivative of the function $B$ with respect to time (frequency).
We can now perform the $t_1$ integral by using 
$\int dt_1 e^{i \omega_1 t_1 } t_1^\alpha = (-1)^\alpha \alpha!/(i \omega_1)^{\alpha} \int dt_1 e^{i \omega_1 t_1 }=2\pi  (-1)^\alpha \alpha ! /(i \omega_1)^{\alpha} \delta(\omega_1)$.
One can further use the identity for derivatives of the $\delta$-function, $\delta^{(n)}(x) = (-1)^n n!/x^n \delta(x)$.
At the same time, we want to move under the frequency integration, $\sum_{n}\int d\omega$, the differentiation away from the $\delta$-function.
The formula can then be written as
\begin{align}
 \begin{split}
 A(t,t') &=  \int \frac{d\omega'}{2\pi} \, d\omega'' \sum_{\substack{n',m',l\\n'',m'',l'}}  \sum_{k=0}^{l} \sum_{k'=0}^{l'} 
 \frac{1}{l! l'!}\frac{1}{2^{l} 2^{l'}} \binom{l}{k} \binom{l'}{k'}\\ 
&\times ~ (-1)^{l+2l'-k'} t^{k} t'^{k'} \delta(\omega'-\omega'' + (m'-n'')\Omega)   \\
 &\times ~ i^{l'-k'} \Big( e^{-i(\omega' + n'\Omega)t}  B_{n'm'}^{(l,0)}(\omega',\bar{t})\Big)^{(0,l'-k')} \\
 &  \times ~ i^{l-k}  \Big(e^{i(\omega'' + m''\Omega)t'} C_{n''m''}^{(l',0)}(\omega'',\bar{t}) \Big)^{(0,l-k)}
 \end{split}
\end{align}
Due to the fact that $-\frac{\Omega}{2} \le \omega < \frac{\Omega}{2}$, the argument of the $\delta$-function above can only be zero if $\omega'=\omega''$ and $m'=n''$.
Hence, the energy integration $\int d\omega'' \sum_n''$ can be straightforwardly performed.
In the following, one wants to use the identity $(fg)^{(n)}=\sum_{k=0}^n\binom{n}{k}f^{(n-k)}g^{(k)}$ and simplify summations.
Eventually, after some steps a relabelling of indices yields the form
\begin{align}
 \begin{split}
  A(t,t') &=  \int \frac{d\omega}{2\pi} \sum_{n,m}  ~ e^{-i(\omega + n\Omega)t} e^{i(\omega + m\Omega)t'}    \\
 &\times ~ \sum_{l,l'} \frac{(-1)^l}{2^{l} 2^{l'}} \frac{i^{l+l'}}{l! l'!}~ \sum_{m'} B_{nm'}^{(l,l')}(\omega,\bar{t}) ~ C_{m'm}^{(l',l)}(\omega,\bar{t})
 \end{split}
\end{align}
According to Eq.~\eqref{Moyal2} we can read off the explicit form of $A_{nm}(\omega,\bar t)$.
After a few more steps of simplification, one finally finds an expression for the Floquet-Moyal product
\begin{equation}
\label{FMP}
A_{nm}(\omega,\bar t) =  e^{-\frac{i}{2} (\partial_{\bar{t}}^{B} \partial_{\omega}^{C} - \partial_{\omega}^{B} \partial_{\bar{t}}^{C})} ~
 \sum_{m'} B_{nm'}(\omega,\bar{t}) ~ C_{m'm}(\omega,\bar{t})
\end{equation}
where $\partial^{B/C}$ is an operator acting only on object $B$ or $C$, respectively.
The Floquet-Moyal expansion is obtained by expanding the exponential fucntion in \ref{FMP}.
One can see that the form of this expression is very similar to the ordinary Moyal product.
Here, however, the formula requires an additional matrix multiplication in the Floquet indices taking care of the fast oscillations.
Note again that this formula only holds in the limit where the inverse Floquet-Wigner transformation \eqref{InversefloquetWigner} is valid.

\section{Semiclassical dynamics and Berry corrections}
\label{appendixberry}
In order to describe the semiclassical dynamics of the Floquet occupations $n_{\nu}(t)$,
i.e., the left-hand side of the Floquet-Boltzmann equation, including leading order Berry phase corrections, we follow the derivation by Wickles and Belzig \cite{Wickles2013}.
Below we will sketch the idea and show why their arguments can be analogously used for the Floquet picture.
To acquire kinetic equations for the individual Floquet states $\nu$, one will additionally have to decouple the QKE by going into the space that diagonalizes the Floquet Hamiltonian.
This unitary transformation is, of course, given by the matrix of instantaneous Floquet eigenstates.
This procedure, however, has a peculiarity: one diagonalizes the objects only after the Floquet-Wigner transformation.
In this form objects are allowed to be functions of two canonical variables, which is (semi-)classically allowed, but quantum mechanically forbidden.
So in order to preserve more of the quantum mechanical nature of our problem we desire a unitary transformation that diagonalizes the object of interest, e.g. $(G^{R})^{-1}$,
already on the level of the initial convolution
\begin{eqnarray}
\label{unitaryFM}
U \circ (G^{R})^{-1} \circ U^\dagger = (\tilde{G}^{R})^{-1}
\end{eqnarray}
with  $U\circ U^\dagger = \mathbbm{1}$ and where '$\circ$' can be viewed as a Floquet-Moyal product.
The obvious problem is that $U$ can generally not be found.
However, one can systematically calculate corrections order by order.
Hence, one writes  $U=U_0 (\mathbbm{1} + U_1 +\cdots)$ with $U_0$ being the unitary matrix that diagonalizes the instantaneous Floquet Hamiltonian, $U_0 H^F U_0^\dagger = \tilde{H}^F$.
Expanding the Floquet-Moyal product yields terms like
\begin{eqnarray}
 U_0 \partial_i U_0^{\dagger} = \partial_i - i \mathcal{A}_i
\end{eqnarray}
where Berry connections have been defined
\begin{eqnarray}
\label{floquetvecpot}
 \mathcal{A}_i \equiv i U_0 (\partial_i U_0^{\dagger})
\end{eqnarray}
Here, $\partial_i$ runs over all possible derivatives (space, momentum, time, energy).
The crucial point to realise here is that $\mathcal{A}_i$ is a matrix in Floquet space.
While in the derivation by Wickles and Belzig this matrix only knew about the original bands of the Hamiltonian, here the object is fully aware of the underlying Floquet structure.
The reason why the Floquet notion is implemented here in such a straight forward manner lies in the fact that it was possible to reduce the fast oscillations to a simple matrix structure.
Hence, the procedure of diagonalization by means of $U$ naturally introduces the Floquet character.

Performing the unitary Floquet-Moyal transformation \eqref{unitaryFM}, keeping only terms up to first order in $\hbar \mathcal A$, and projecting the expression onto the eigenstates of the problem,
i.e., here the Floquet states, the object $(\tilde{G}^{R})^{-1}$ can be approximated as
\begin{eqnarray}
 (\tilde{G}^{R})^{-1} = &&(\tilde{G}^{R})^{-1}_{0} - \frac{\hbar}{2} \{ \mathcal A_{\boldsymbol{\pi}}^{(d)} ,\partial_{\vec x} (\tilde{G}^{R})^{-1}_{0} \} \\
&&~~ + \frac{\hbar}{2} \{ \mathcal A_{\vec x} ,\partial_{\boldsymbol{\pi}}^{(d)} (\tilde{G}^{R})^{-1}_{0} \} + \mathcal O((\hbar\mathcal A)^2) \nonumber
\end{eqnarray}
where $\vec x = (t,\vec r)$, $\boldsymbol{\pi}=(\omega,-\vec p)$, $\mathcal A^{(d)} = \sum_i \mathcal P_i A \mathcal P_i$, with $\mathcal P_i$ being a projector, 
and $(\tilde{G}^{R})^{-1}_{0} =U_0 (G^{R})^{-1} U_0^{\dagger}$.
One can observe that the expression above is nothing but a Taylor expansion to first order.
Hence, after introducing band projected kinetic variables $\vec X =\vec x - \hbar \mathcal A_{\boldsymbol{\pi}}^{(d)}$ and $\boldsymbol{\Pi}=\boldsymbol{\pi} + \hbar A^{(d)}_{\vec x}$, 
one can write
\begin{eqnarray}
 (\tilde{G}^{R})^{-1} \approx (\tilde{G}^{R})^{-1}_{0}(\vec X,\boldsymbol{\Pi})
\end{eqnarray}
So we can indeed express the object in its trivially diagonalized form.
However, the prize we need to pay is the change from canonical to kinetic variables.
Thus, the derivatives appearing in the Moyal product have to be changed accordingly, and the new expression reads ($\hbar =1$)
\begin{eqnarray}
 \circ \to \exp[-\frac{i}{2} (\partial_{\vec X}^{B} \partial_{\boldsymbol{\Pi}}^{C}-\partial_{\boldsymbol{\Pi}}^{B}\partial_{\vec X}^{C})
 +\frac{i}{2}\partial_{\mu}^{B} \hat{\Omega}_{\mu\nu} \partial_{\nu}^{C}]
\end{eqnarray}
for a product of $B \circ C$ where the partial derivative $\partial^{B/C}$ acts only on object $B$ or $C$, respectively.
Here, $\nu,\mu=0,\cdots,6$ run over all time, position and momentum indices.
The Floquet-Moyal product '$\circ$' also requires a multiplication of Floquet matrices $B$ and $C$.
Hence, the entries of the object $\hat{\Omega}$, the  \textit{Berry-curvature tensor}, are in fact matrices in Floquet space.
The explicit form (for non-crossing Floquet bands) reads
\begin{eqnarray}
 \hat{\Omega}_{\mu\nu} = \frac{\partial \mathcal A_{\nu}^{(d)} }{\partial \mu } - \frac{\partial A_{\mu}^{(d)}}{\partial \nu }
\end{eqnarray}
Using Eq.~\eqref{floquetvecpot} and the fact that $U_0$ is simply the matrix of Floquet eigenstates one can write down the Berry-curvature tensor for the Floquet state $\xi$
\begin{eqnarray}
\Omega^\xi_{\mu \nu}&=&\sum_{i,n} (\partial_\nu \bar{\phi}_{t,\vec k, \xi}^n(i)) (\partial_\mu \phi_{t,\vec k, \xi}^n(i)) \\
&&\qquad \qquad -~(\partial_\mu \bar{\phi}_{t,\vec k, \xi}^n(i)) (\partial_\nu \phi_{t,\vec k, \xi}^n(i)) \nonumber
\end{eqnarray}

Note that the projection onto on-shell processes performed on the right hand side of the Boltzmann equation is already encoded in the procedure above by virtue of Eq.~\eqref{unitaryFM}.
At the same time, the instruction to go from canonical to kinetic variables is obsolete for the right hand side, since there we allow only zero order terms to contribute (for which $\vec X = x$ and $\vec \Pi = \pi$).

\begin{table}[t!]
\begin{tabular}{c||c|c||c|c}
\hline\hline
$j$ & $J_j/\hbar\, (\rm{Hz})$ & $\vec v_j/\lambda$ & $J_j^A/\hbar\, (\rm{Hz})$ &  $\vec u_j/\lambda$ \\ \hline
0 & $- 2\pi \cdot 746$  & $(0.438, 0)$ & - & - \\ 
1 & $- 2\pi \cdot 527$ & $(-0.062, 0.5)$ & $ 2\pi \cdot 14$ & $(0.5, -0.5)$ \\ 
2 & $- 2\pi \cdot 527$ & $(-0.062, -0.5)$ & $ 2\pi \cdot 14$ & $(-0.5, -0.5)$ \\ 
3 & $- 2\pi \cdot 126$ & $(-0.562, 0)$ & $ 2\pi \cdot 61$ & $(0,1.0)$ \\
\hline\hline
\end{tabular}
\caption{Summary of all hopping strengths and associated lattice vectors as used in Appendix~\ref{alternativepara} as well as in \cite{Jotzu2014}.
While $J_j$ describes the hopping between nearest neighbours with $\vec v_j$ connecting these points (from sublattice $\mathcal A$ to $\mathcal B$),
$J_j^A$ describes the hopping between next-nearest neighbours with $\vec u_j$ connecting points on the same sublattice. 
Note also that $\vec e_1 =(1,0)$ and $\vec e_2 =(0,1)$, and that $J_j^A =J_j^B$.
}
\label{tab_hoppings}
\end{table}

\section{Alternative set of parameters for the interacting Haldane model}
\label{alternativepara}

\begin{figure}[t!]
   \centering
   \includegraphics{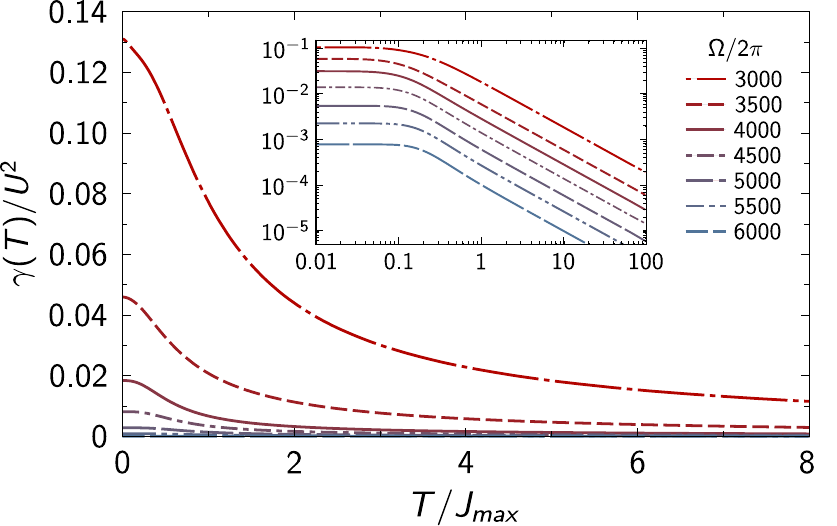}
   \caption{(Color online) Heating rate per lattice site (in units of $1/U^2$) plotted against dimensionless temperature $T/J_{\rm max}$, 
   where $J_{\rm max}$ is the maximal hopping amplitude of the anisotropic Hubbard model at hand.
   Different curves describe different driving frequencies $\Omega$ ranging from $4 |J_{\rm max}|\approx2 \pi\cdot 3000$\,Hz to $8 |J_{\rm max}|\approx2 \pi\cdot 6000$\,Hz, see legend.
   The inset shows a double logarithmic plot.
   The solid curve ($\Omega  = 2 \pi \cdot 1080\,\rm Hz \approx 5.4 |J_{\rm max}|$) is the frequency used in the main text of \cite{Jotzu2014}.}
 \label{g_vs_T_A}
\end{figure}

\begin{figure}[t!]
   \centering
   \includegraphics{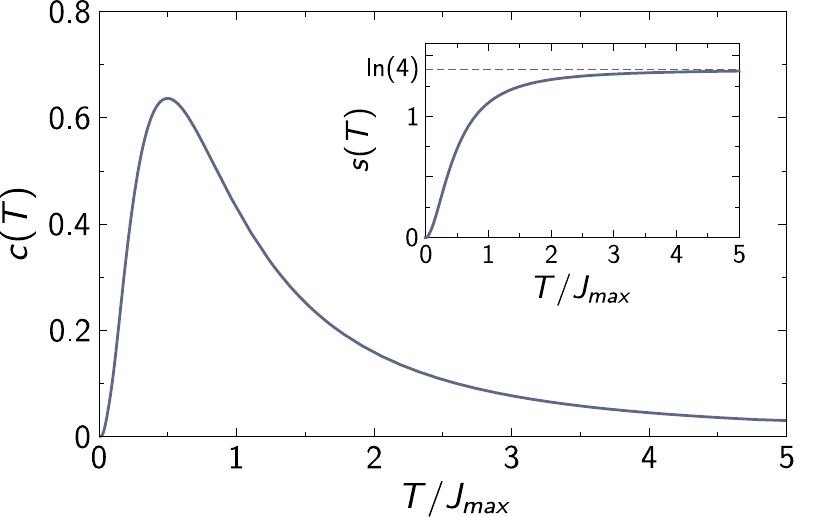}
   \caption{(Color online) Specific heat per lattice site, $c(T)$, 
   plotted as function of dimensionless temperature $T/J_{\rm max}$ for a driving frequency $\Omega =2 \pi \cdot 4000 \rm{Hz} \approx 5.4 |J_{\rm max}|$.
   The inset shows the corresponding entropy per lattice site $(k_B = 1)$ as function of $T/J_{\rm max}$. }
 \label{c_vs_T_A}
\end{figure}

\begin{figure}[t!]
   \centering
   \includegraphics{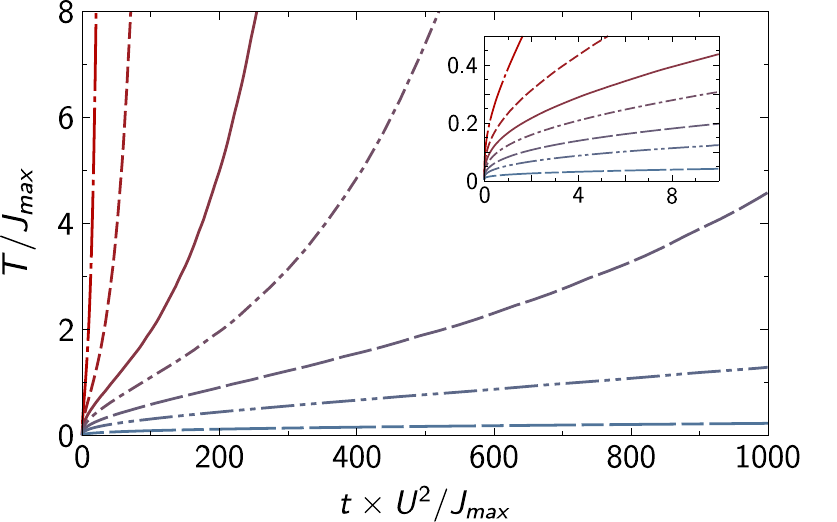}
   \caption{(Color online) Dimensionless temperature, $T/J_{\rm max}$, of the interacting system as a function of time for different driving frequencies $\Omega$ 
   ($\Omega=2 \pi \times 3000, \dots, 6000$\,Hz), 
   see legend of Fig.~\ref{g_vs_T_A}.  
   The inset clarifies that the curves undergo two distinct regimes before rising with exponential speed. 
   }
 \label{T_vs_t_A}
\end{figure}

\begin{figure}[t!]
   \centering
   \includegraphics{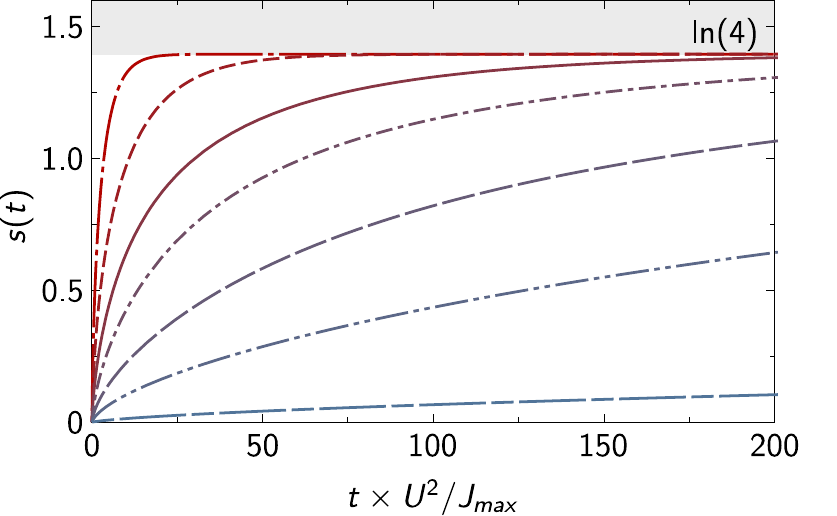}
   \caption{(Color online) Entropy per lattice site (per particle) as a function of time for different driving frequencies $\Omega$ ($\Omega=2 \pi \times 3000, \dots, 6000$\,Hz), 
   see legend of Fig.~\ref{g_vs_T_A}. 
}
 \label{s_vs_t_A}
\end{figure}

In Sec.~\ref{sec:Haldane} we presented the predictions of our Floquet-Boltzmann equation \ref{FBE} for the recently realized Haldane model \cite{Jotzu2014} for an isotropic setup,
characterized by a single nearest-neighbour hopping amplitude $J$ only.

Here we repeat the analysis for the (spinfull) model described in the main text of \cite{Jotzu2014}.
The geometry of the model remains the same as described in Sec.~\ref{sec:Haldane}, see  Fig.~\ref{lattice},
but hoppings are anisotropic and also  next-nearest neighbour hopping amplitudes $J^A_j$ are taken into account.
A summary of the experimental parameters  of Ref.~\cite{Jotzu2014} which we also used in our calculations can be found in Tab.~\ref{tab_hoppings}.

We then apply the exact same arguments and procedures as in Sec.~\ref{sec:Haldane}.
Note that all approximations and assumptions made above also hold here because the driving frequencies is here now tuned to higher absolute values $\Omega=2\pi\cdot 3000,\dots, 6000\,\rm Hz$.
Translating these frequncies into units of the maximal tunnelling amplitude $J_{\rm max} = J_{0}$, 
one sees that the different values of $\Omega$ span the same regime as above, i.e., $\Omega \approx 4|J_{\rm max}|,\dots,8|J_{\rm max}|$.
We now recalculate Figs.~\ref{g_vs_T}-\ref{s_vs_t} for the alternative set of parameters presented here and plot the results in Figs.\ref{g_vs_T_A}-\ref{s_vs_t_A}.
Note that we used $J_{\rm max}$ throughout to rescale energies.
In fact, comparing the two sets of parameters clearly reveals that the results are qualitatively unaffected.
Hence, all conclusions drawn from the results in Sec.~\ref{sec:Haldane} also apply here.

\bibliographystyle{aipnum4-1}
\bibliography{bib}

\end{document}